\def\be{\begin{equation}}
\def\eq{\end{equation}}

\documentclass[showpacs,preprintnumbers,amsmath,amssymb]{revtex4}

\usepackage{graphicx}
\usepackage{dcolumn}
\usepackage{bm}
\usepackage{hyperref}

\begin{document}

\preprint{EFI 07-07}
\preprint{hep-ph/0703177}

\title{Masses and Mixings in a Grand Unified Toy Model}
\author{David McKeen}
\email{mckeen@theory.uchicago.edu}
\author{Jonathan L. Rosner}
\email{rosner@hep.uchicago.edu}
\author{Arun M. Thalapillil}
\email{madhav@theory.uchicago.edu}
\affiliation{Enrico Fermi Institute and Department of Physics, University of Chicago, 5640 South Ellis Avenue, Chicago, IL 60637}
\date{\today}

\begin{abstract}
{The generation of the fermion mass hierarchy in the standard model of particle physics is a long-standing puzzle. The recent discoveries from neutrino physics suggests that the mixing in the lepton sector is large compared to the quark mixings. To understand this asymmetry between the quark and lepton mixings is an important aim for particle physics. In this regard, two promising approaches from the theoretical side are grand unified theories and family symmetries. In this paper we try to understand certain general features of grand unified theories with Abelian family symmetries by taking the simplest $SU(5)$ grand unified theory as a prototype. We construct an $SU(5)$ toy model with $U(1)_{F}\otimes\mathbb{Z}^{'}_{2}\otimes\mathbb{Z}^{''}_{2}\otimes\mathbb{Z}^{'''}_{2}$  family symmetry that, in a natural way, duplicates the observed mass hierarchy and mixing matrices to lowest approximation. The system for generating the mass hierarchy is through a Froggatt-Nielsen type mechanism. One idea that we use in the model is that the quark and charged lepton sectors are hierarchical with small mixing angles while the light neutrino sector is democratic with larger mixing angles. We also discuss some of the difficulties in incorporating finer details into the model without making further assumptions or adding a large scalar sector. }
\end{abstract}
\pacs{12.15.Ff, 14.60.Pq, 12.10.Dm, 11.30.Hv}

\maketitle


\begin{section}{Introduction}
In the Standard Model (SM) the Yukawa coupling constants can be freely adjusted without disturbing the internal consistency of the theory; one must rely on experiments to fix their values. They are now tightly constrained and it is interesting to try to predict the mass ratios and mixing angles from some first principle calculation, incorporating physics that is beyond the SM. Among the attempts to explain and relate some of the arbitrary parameters in the SM are grand unified theories (GUTs) and family symmetries. 
\par
GUTs embed  the SM group $SU(3)_{C}\times SU(2)_{L}\times U(1)_{Y}$ into a larger group $G$ and, as a result, previously independent SM parameters may become related (see, for example, \cite{pala}). One of the earliest and most interesting attempts in this direction was the $SU(5)$ theory of Georgi and Glashow \cite{gegla}. Among other things, in the `minimal' version, it predicted that at the GUT scale the charged leptons and down quarks in each generation have equal masses. On the other hand family symmetries between the three fermion generations generally lead to inter-family mass relations. Froggatt and Nielsen (FN) long ago suggested that a spontaneously broken $U(1)$ family symmetry may be the cause for the observed hierarchy of fermion masses \cite{frni}. This is because in the FN mechanism the Yukawa coupling constants are derived naturally when scalar fields, transforming under the family symmetry group, attain a vacuum expectation value (VEV), spontaneously breaking the symmetry. It must be noted that in most cases while the GUTs relate the quark and lepton masses in the \textit{same} generation, the family symmetry relates the masses of fermions in \textit{different} generations. Thus, sometimes we refer to them as `vertical' and `horizontal' symmetries respectively. There have been some pioneering studies in recent years using FN to understand quark and lepton mass hierarchies \cite{fnm}. 
\par
 The main aim of this paper is to explore some general considerations in GUT models with Abelian family symmetries taking the $SU(5)$ GUT as an example. In this paper we explore the generation of mass hierarchies in the quark and lepton sectors in GUTs through the spontaneous breaking of an Abelian family symmetry. The attempt will be to understand general features of GUT models with the simplest family symmetries and the least number of assumptions. We also require that the scalar sector is minimal. In this regard, as previously mentioned, we consider the simplest GUT group, $SU(5)$,  as the prototype and take the family symmetry to be $U(1)_{F}\otimes\mathbb{Z}^{'}_{2}\otimes\mathbb{Z}^{''}_{2}\otimes\mathbb{Z}^{'''}_{2}$. One reason for the choice apart from its simplicity is that the Abelian groups are universally present in many models where a higher GUT group (like $E_{6}$ or $SO(10)$) undergoes symmetry breaking. A toy model is constructed which captures the general features of the quark and lepton mass hierarchies. It also naturally incorporates the qualitative features of the quark and lepton mixing matrices to lowest approximation. Other models in the literature aimed at understanding the mixing patterns have been constructed based on, for example, $A_{4}$ \cite{a4} and $SU(3)$ \cite{su3}.

 \par
 We attempt to explain the differences between the quark mixing matrix and the lepton mixing matrix as a consequence of the presence of right-handed Majorana neutrinos. There may be two scenarios to create the difference in the quark and lepton mixings. The quark and charged lepton Yukawa matrices may be hierarchical while the light neutrino matrix may be non-hierarchical. By a hierarchical Yukawa matrix we mean that the mixing angles in the diagonalizing matrices are small. A sufficient condition for this is that the Yukawa matrix is very close to a diagonal matrix with the off-diagonal elements relatively small. The non-hierarchical Yukawa matrix on the other hand has larger mixing angles and is far from being diagonal. This would lead the quark and charged lepton diagonalizing matrices to be close to the unit matrix and the neutrino diagonalizing matrix to be tri-bimaximal \cite{hps}. The other scenario is that the light neutrino matrix may be hierarchical while the quark and charged lepton Yukawa matrices are non-hierarchical. In this case the neutrino diagonalizing matrix will be close to unity while the other three matrices will be close to tri-bimaximal \cite{hps}. The essential point is that in both the above cases the difference in the quark and lepton mixings is explained through a mismatch of one matrix with the other three matrices. For an interesting alternative the reader is referred to \cite{mpg}. In our model both the matrices that contribute to the Cabibbo-Kobayashi-Maskawa (CKM) matrix \cite{km} are themselves close to a unit matrix. In the lepton sector the matrix that diagonalizes the charged lepton Yukawa matrix is also close to a diagonal unit matrix, but the contribution from the neutrino sector is close to a tri-bimaximal matrix. Thus the Pontecorvo-Maki-Nakagawa-Sakata (PMNS) matrix \cite{mns}  comes out being very close to a tri-bimaximal matrix \cite{hps}, so in the model we construct the difference between the CKM and PMNS matrices is solely due to the neutrino sector.
 \par
The observational constraints on the quark and lepton masses and mixing angles are reviewed in section II.  In section III we give a short description of  the $SU(5)$ GUT and the idea behind Abelian family symmetries.  Then in section IV we construct the toy model that captures to zeroth approximation the features in the fermion mass hierarchy and electroweak mixing matrices. We go on to discuss in section V some of the difficulties in the GUT models when one proceeds to incorporate finer details in the mass hierarchy and mixing angles, again by taking the $SU(5)$ as our prototype GUT. Section VI is the conclusion.
  \end{section}


\begin{section}{Experimental Constraints}

\par
The pattern of quark and lepton masses may provide us with a rare insight into physics beyond the standard model. Unfortunately, determination of the mass eigenvalues and mixing angles is not sufficient to determine the complete structure of the Yukawa matrices. Thus one is usually led to make some assumptions about the Yukawa matrices themselves. To motivate our discussions in the rest of the paper we briefly review the conditions that any model of masses and mixings must satisfy.

\par
The observed fermion mass hierarchy is most apparent in the quark sector. Since the mass eigenstates have unique charges, we may consider the charge $+\frac{2}{3}$ and $-\frac{1}{3}$ quarks separately. The masses of the $+\frac{2}{3}$ charged `up' quarks are \cite{{pdg,wang:moriond}}
\be
\begin{split}
 m_{u}&\simeq1.5-3~{\text MeV}~~~,\\
 m_{c}&\simeq1.16-1.34~{\text GeV}~~~,\\
 m_{t}&\simeq169.1-172.7~{\text GeV}~~~,
\label{upqmasses}
\end{split}
\eq
and the masses of the $-\frac{1}{3}$ charged down quarks are
\be
\begin{split}
m_{d}&\simeq3-7~{\text MeV}~~~,\\
m_{s}&\simeq70-120~{\text MeV}~~~,\\
m_{b}&\simeq4.1-4.2~{\text GeV}~~~.
\end{split}
\label{downqmasses}
\eq
It is to be noted that all the quark masses except the top quark mass are in the $\overline{\text{MS}}$ scheme \cite{pdg}. In this scheme, for the light quarks $u$, $d$ and $s$ the renormalization scale is taken to be about $2~\text{GeV}$ and for the $c$, $b$, and $t$ quarks the scale is taken to be at the threshold for pair production.  One observes here that there is a strong mass hierarchy among the three generations in both the up and down quark sectors with the mass spacings in the up sector larger than those in the down sector. There is some ambiguity in the measurement of the absolute quark masses since they are scheme dependent, but the ratios of the masses are more concrete. The light quark mass ratios are measured to be \cite{pdg}
\be
\begin{split}
&\frac{m_{u}}{m_{d}}~=~0.3-0.6~~~, \\
&\frac{m_{s}}{m_{d}}~=~17-22~~~,\\
&\frac{m_{s}-(m_{u}+m_{d})/2}{m_{d}-m_{u}}~=~30-50~~~.
\end{split}
\label{udqmratios}
\eq

\par
The mixing of the electroweak eigenstates with the mass eigenstates in the case of the quarks is parametrized by the CKM matrix \cite{km}. Recent precision measurements have greatly improved knowledge of the CKM matrix parameters. The experimental constraints on the CKM parameters \cite{{pdg},{czm}} are

\be
\Bigl\lvert\mathcal{U}^{Exp.}_{CKM}\Bigr\rvert\simeq
\begin{pmatrix}
0.974&0.226-0.228& 0.004 \\
0.226-0.228&0.973&0.041-0.042 \\
0.008&0.041-0.042& 0.999
\end{pmatrix}~~~
\label{ckmobs}
\eq
where only the magnitudes of the elements are shown and the Dirac CP phase \cite{pdg} is not explicitly included. The CKM matrix to first approximation is observed to be very close to a unit matrix. All experiments to date strongly suggest that unitarity is preserved. Using the standard parametrization of  the CKM matrix \cite{pdg} in terms of the angles $\theta^{q}_{23}$, $\theta^{q}_{13}$ and $\theta^{q}_{12}$ the above may be interpreted as a constraint on the angles
\be
\begin{split}
2.37^{\circ}\leq~&\theta^{q}_{23}\leq2.43^{\circ}~~~,\\
0.222^{\circ}\leq~&\theta^{q}_{13}\leq0.232^{\circ}~~~,\\
13.07^{\circ}\leq~&\theta^{q}_{12}\leq13.19^{\circ}
\label{qangles}~~~.
\end{split}
\eq
Since the CKM matrix is nearly an identity matrix the mixing angles are rather small. This nature of the CKM matrix strongly suggests that possibly the quark sector Yukawa matrices are of a `hierarchical' structure. As mentioned before, by hierarchical we mean that the mixing angles that appear in the diagonalizing matrices are small. This will be a guiding principle that we will adopt in the paper to construct the toy model. In fact we assume a stronger condition that the Yukawa matrices for the quarks are very close to diagonal matrices. This is a sufficient condition for small mixing angles in the quark sector as well as in the charged lepton sector due to the properties of the $SU(5)$ group. 
\par
The masses of the charged leptons have been measured much more unambiguously than the quark masses. The charged lepton sector is also seen to exhibit a large mass hierarchy. Their masses are measured to be \cite{pdg}
\be
\begin{split}
m_{e}~&\simeq~0.511~{\text MeV}~~~,\\
m_{\mu}~&\simeq~105.7~{\text MeV}~~~,\\
m_{\tau}~&\simeq~1777~{\text MeV}~~~.
\end{split}
\eq
The $e$, $\mu$ and $\tau$ masses are the \textit{pole masses}~\cite{pdg}. Note that this mass hierarchy is more similar to the charge $-\frac{1}{3}$ quark sector than the charge $+\frac{2}{3}$ quark sector. This observation has been the basis for many schemes in which the down quark and charged lepton masses unify at some energy scale \cite{pala}.

We now turn to the neutrino sector. Observations in the neutrino sector currently provide the strongest indication for physics beyond the standard model. Neutrino oscillations and the question of whether neutrinos are Dirac or Majorana have spurred progress in particle theory and experiments. The experimental constraints on the neutrino parameters from neutrino oscillation experiments are (see for example \cite{strumvis} and references therein)
\be
\begin{split}
&\Delta m^{2}_{32}~\simeq~2.5\times10^{-3}~~{\text eV}^{2}~~~,\\
&\Delta m^{2}_{21}~\simeq~8.1\times10^{-5}~~{\text eV}^{2}~~~,
\end{split}
\label{numsq}
\eq
where $\Delta m^{2}_{ij}=m^{2}_{\nu_{i}}-m^{2}_{\nu_{j}}$, $\theta^{\nu}_{23}$ is the atmospheric angle and $\theta^{\nu}_{12}$ is the solar angle. The above result for the mass squared differences suggest that at least two of the neutrinos have non-zero masses. Apart from the above constraints we also have 
\be
\sum_{i}m_{\nu_{i}}~\lesssim~0.6~~{\text eV}
\label{numcos}
\eq
from cosmological considerations \cite{strumvis}. Results from the LSND \cite{lsnd} experiment seemed to favor the addition of one or more sterile neutrinos, but this observation has not been confirmed by the MiniBooNE experiment \cite{jcwl:miniboone}. For the considerations of our study we assume that there are only three families of neutrinos. 
As with the CKM matrix in the quark sector, the mixing in the lepton sector is described by the PMNS matrix \cite{mns}. The experimental bounds on the PMNS matrix elements give \cite { {pdg}, {strumvis}, {mnsexp}}
\be
\Bigl\lvert\mathcal{U}^{~Exp.}_{PMNS}\Bigr\rvert~\simeq~
\begin{pmatrix}
0.79-0.86 &0.50-0.61& 0-0.16 \\
0.24-0.52&0.44-0.69&0.63-0.79 \\
0.26-0.52&0.47-0.71& 0.60-0.77
\end{pmatrix}~~~.
\label{exppmns}
\eq
The most striking difference we immediately observe in the PMNS matrix, as compared to the CKM matrix, is its large deviation from an identity matrix. When parametrized by the angles $\theta^{\nu}_{23}$, $\theta^{\nu}_{13}$ and $\theta^{\nu}_{12} $ the above observation is converted to the bounds on the angles

\be
\begin{split}
36^{\circ}\leq~&\theta^{\nu}_{23}\leq54^{\circ}~~~,\\
0^{\circ}\leq~&\theta^{\nu}_{13}\leq10^{\circ}~~~,\\
30^{\circ}\leq~&\theta^{\nu}_{12}\leq38^{\circ}
\label{nuangles}~~~.
\end{split}
\eq
Thus the mixing angles are large in the lepton sector compared to the quark sector. This intuitively suggests that the Yukawa matrix contributing to PMNS is of a non-hierarchical or `democratic' nature. By this we mean that the Yukawa matrix elements are all of roughly the same order and the mixing angles are therefore large. Thus one possible way that the CKM and PMNS matrices may be made to come out differently is through a mismatch in the rotation matrices that contributes to each of them. But we still have the freedom to choose which of the matrices are hierarchical and which are democratic. In the model we study it is found that it is more natural to implement hierarchy in the quark sector (and thus the charged lepton sector) and make the light neutrino sector democratic. This idea will be elaborated and implemented in section IV.
\par
Although the absolute masses of the fermions run through renormalization group evolution, the intra-family mass ratios themselves do not run significantly. This may be readily seen from the well-known expressions (see for example \cite{chenli})
\be
\begin{split}
&\frac{m_{u}(q)}{m_{u}(q_{0})}~=~\left[\frac{g_{1}(q)}{g_{1}(q_{0})}\right]^{-\frac{6}{10n_{f}}}\left[\frac{g_{3}(q)}{g_{3}(q_{0})}\right]^{\frac{8}{11-2n_{f}/3}}~~~,\\
&\frac{m_{d}(q)}{m_{d}(q_{0})}~=~\left[\frac{g_{1}(q)}{g_{1}(q_{0})}\right]^{\frac{3}{10n_{f}}}\left[\frac{g_{3}(q)}{g_{3}(q_{0})}\right]^{\frac{8}{11-2n_{f}/3}}~~~,\\
&\frac{m_{e}(q)}{m_{e}(q_{0})}~=~ \left[\frac{g_{1}(q)}{g_{1}(q_{0})}\right]^{-\frac{27}{10n_{f}}}~~~.
\end{split}
\eq
In the above equations $q$ and $q_{0}$ are energy scales, $n_{f}$ is the number of quark flavors  and $g_{1}(q)$, $g_{2}(q)$ and $g_{3}(q)$ are the coupling constants corresponding to $U(1)_{Y}$, $SU(2)_{L}$ and $SU(3)_{C}$ respectively at the energy scale $q$. The subscripts $u$, $d$ and $e$ stand for any up sector quark, down sector quark and charged lepton respectively. Using the scale dependence of the coupling constants, we find central values of the fermion masses at a unification scale of $10^{16}~\text{GeV}$ to be
\begin{align}
&&&&&
&m_{u}&\simeq0.57~{\text MeV}~,&m_{c}&\simeq0.328~{\text GeV}~,&m_{t}&\simeq71.1~{\text GeV}~~~,
&&&&&&\nonumber\\
&&&&&
&m_{d}&\simeq1.31~{\text MeV}~,&m_{s}&\simeq24.8~{\text MeV}~,&m_{b}&\simeq1.30~{\text GeV}~~~,
&&&&&&\\
&&&&&
&m_{e}&\simeq0.458~{\text MeV}~,&m_{\mu}&\simeq95.8~{\text MeV}~,&m_{\tau}&\simeq1.62~{\text GeV}~~~.
&&&&&&\nonumber
\end{align}
Based on the above experimental constraints we are led to assign approximate phenomenological constraints on the mass ratios at the GUT scale to be
\begin{align}
&&&&&&&
&\frac{m_{u}}{m_{t}}&\simeq\lambda^{8.0}\thicksim\lambda^{8}~,
&\frac{m_{c}}{m_{t}}&\simeq\lambda^{3.7}\thicksim\lambda^{4}~,
&&&&&&&&\nonumber\\
&&&&&&&
&\frac{m_{d}}{m_{b}}&\simeq\lambda^{4.7}\thicksim\lambda^{4}~,
&\frac{m_{s}}{m_{b}}&\simeq\lambda^{2.7}\thicksim\lambda^{2}~,
&&&&&&&&\label{qrel}\\
&&&&&&&
&\frac{m_{e}}{m_{\tau}}&\simeq\lambda^{5.6}\thicksim\lambda^{5}~,
&\frac{m_{\mu}}{m_{\tau}}&\simeq\lambda^{1.92}\thicksim\lambda^{2}~,
&&&&&&&&\nonumber\\
&&&&&&&
&\frac{m_{b}}{m_{t}}&\simeq\lambda^{2.7}\thicksim\lambda^{3}\nonumber~,
&\frac{m_{b}}{m_{\tau}}&\simeq\lambda^{0.15}\thicksim 1~,
&&&&&&&&\nonumber
\end{align}
where we have parametrized the mass ratios by $\lambda$, the Cabibbo angle ($\lambda\simeq 0.23$). To round off the exponents in the above ratios we balanced the desire to have rough equality of the charged lepton and down quark masses as required by a minimal $SU(5)$ theory~ \cite{gegla}, to have equal logarithmic mass spacing between the generations in both the up and down quark sectors, and to have the logarithmic mass spacing in the up quark sector be twice that in down quark sector. As we shall see later this will lead to a desirable light neutrino Yukawa matrix due to the relations in the $SU(5)$ GUT. The set of approximate ratios above will be another of our guiding principles in constructing the toy model.
\par
From Eqs.\ (\ref{ckmobs}) and (\ref{exppmns}) the mixing matrices to first approximation may be written as an identity matrix for the quark sector
\be
\Bigl\lvert\mathcal{U}_{CKM}\Bigr\rvert~\thicksim~
\begin{pmatrix}
1 &0& 0 \\
0&1&0 \\
0 &0& 1
\end{pmatrix}~~~,
\label{ckm}
\eq
and as a tri-bimaximal matrix for the lepton sector
\be
\Bigl\lvert\mathcal{U}_{PMNS}\Bigr\rvert~\thicksim~
\begin{pmatrix}
\frac{2}{\sqrt{6}} &\frac{1}{\sqrt{3}} & 0 \\
\frac{1}{\sqrt{6}} &\frac{1}{\sqrt{3}}&\frac{1}{\sqrt{2}} \\
\frac{1}{\sqrt{6}} &\frac{1}{\sqrt{3}}& \frac{1}{\sqrt{2}}
\end{pmatrix}
\label{pmns}
\eq
where, again, CP-phases and signs are not explicitly included.
We assume that at the GUT scale the forms of these mixing matrices are essentially unchanged and mirror their form at the electroweak scale. This is an \textit{assumption} in the construction of the toy model. For the CKM matrix, it has been noted in the literature that at the GUT scale the mixing angles in most scenarios do not deviate much from their electroweak values \cite{babu}. There have been many investigations into the renormalization group running of the PMNS matrix to reconcile the differences in the quark and lepton mixing angles \cite{{mpg}, {rgmix}}. It must also be pointed out that the assumption of the mixing matrix texture at the GUT scale being essentially similar to the structure at the electroweak scale depends on various parameters, for example the Majorana phases in the PMNS matrix and the neutrino mass hierarchy \cite{rgmix}. In the toy model we assume the form for the mixing matrices to be as in Eqs.\ (\ref{ckm}) and (\ref{pmns}). We now proceed to give a very brief introduction to the $SU(5)$ GUT and FN mechanism from Abelian family symmetries.
 \end{section}
 

 \begin{section}{The SU(5) GUT and Abelian Family Symmetries}
\par
In the $SU(5)$ GUT, the usual SM fermions and the right-handed neutrino are accomodated in the representations (see for example \cite{{gegla},{slansky}})
\begin{align}
&&&&&&&
&\Psi_{a}~&:&\bm{ 5}^*&\rightarrow (\bm{ 3}^*, \bm{ 1})_{\frac{1}{3}}\oplus(\bm{ 1},\bm{ 2})_{-\frac{1}{2}}~~~,\nonumber
&&&&&&\\
&&&&&&&
&\Psi^{ab}~&:&\bm{ 10}&\rightarrow (\bm{ 3}^*, \bm{ 1})_{-\frac{2}{3}}\oplus(\bm{ 3},\bm{ 2})_{\frac{1}{6}}\oplus(\bm{ 1},\bm{ 1})_{1}~~~,
\label{su5reps}
&&&&&&\\
&&&&&&&
&N^{R}~&:&\bm{ 1}~&\rightarrow (\bm{ 1}, \bm{ 1})_{0}\nonumber
&&&&&&
\end{align}
where the branching rules are in terms of the representations of $SU(3)_{C}\times SU(2)_{L}\times U(1)_{Y}$ of the SM. That is, for a given family, the $\bm{1}$ contains the right-handed neutrino $N^{R}$, the $\bm{ 5}^*$ contains $\bar{d}_R$ and $L=(\nu_L~e_L)$ while the $\bm{10}$ contains $Q=(u_L~d_L)$, $\bar{u}_R$, and $\bar{e}_R$. For each generation the representations are replicated with the appropriate fields. More explicitly the fermions in the first generation, for example, are accommodated as

\begin{align}
\Psi_{a}~&:~~~~~~
\begin{pmatrix}
d_{r}^{c} &d_{g}^{c}&d_{b}^{c}&e^{-}&-\nu_{e} \\
\end{pmatrix}~~~,\nonumber
\\
\nonumber\\
\Psi^{ab}~&:~
\begin{pmatrix}
0 &u_{b}^{c}& -u_{g}^{c}&u_{r}&d_{r} \\
 -u_{b}^{c}&0&u_{r}^{c}&u_{g}&d_{g}  \\
 u_{g}^{c}&-u_{r}^{c}&0&u_{b}&d_{b}\\
 -u_{r}&-u_{g}&-u_{b}&0&e^{+}\\
 -d_{r}&-d_{g}&-d_{b}&-e^{+}&0\\
\end{pmatrix}
\label{su5reps:exp}~~~,
\\
\nonumber\\
N^{R}~&:~~~~~~~~~~~~~~~~~N^{R}_{1}\nonumber
\end{align}
where $r$, $g$, and $b$ are color indices and the subscript on the right-handed neutrino indicates the generation. All the fields in the $\bm{10}$  and $\bm{ 5}^*$ are left-handed. Using Young tableaux one may calculate the direct product decompositions to be
\be
\begin{split}
\bm{5}^{*}\otimes\bm{5}^{*}&=\bm{15}\oplus\bm{10}^{*}~~~,\\
\bm{5}^{*}\otimes\bm{10}&=\bm{45}\oplus\bm{5}~~~,\\
\bm{10}\otimes\bm{10}&=\bm{50}\oplus\bm{45}^{*}\oplus\bm{5}^{*}~~~.
\end{split}
\label{fivten}
\eq
This determines the Higgs scalars that are required to write the Yukawa terms that would then generate the fermion masses. We note that unlike the  SM the $SU(5)$ symmetry introduces some restrictions when one tries to  write down invariant Yukawa terms.  It is seen from the above direct products that the Yukawa matrix for the up quarks comes from the product of the ten-dimensional representations: $\bm{ 10}_i\otimes\bm{ 10}_j$ where $i$ and $j$ label the generation.  The down quark Yukawa matrix results from $\bm{ 10}_i\otimes\bm{ 5}_j^*$ and that of the charged leptons from $\bm{ 5}_i^*\otimes\bm{ 10}_j$.  One sees therefore that in a minimal $SU(5)$ model the charged leptons and down quarks have equal masses. The Higgs sector in `minimal' $SU(5)$ consists of just a Higgs in the $\bm{5}$ representation \cite{gegla}. The down quark and charged lepton mass relations are improved if the Higgs sector is extended to include a $\bm{45}$ \cite{geja}. The branching rules for the Higgs are \cite{slansky}

\be
\bm{5}_{H}~\rightarrow~(\bm{3}, \bm{1})\oplus(\bm{1}, \bm{2})
\label{fivhig}
\eq
and
\be
\bm{45}_{H}~\rightarrow~(\bm{3}, \bm{1})\oplus(\bm{1}, \bm{2})\oplus(\bm{3}, \bm{3})\oplus(\bm{3}^{*}, \bm{1})\oplus(\bm{3}^{*}, \bm{2})\oplus(\bm{6}^{*}, \bm{1})\oplus(\bm{8}, \bm{2})
\label{ffivhig}~~~.
\eq

\par
The hierarchical nature of the fermion masses across the three generations strongly suggests the possibility of a spontaneously broken family symmetry. The presence of such a family symmetry may then naturally lead to the observed masses and mixings via nonrenormalizable terms in the low-energy effective Lagrangian. In the original paper by Froggatt and Nielsen the family symmetry was assumed to be $U(1)$ \cite{frni}. In this study we take the horizontal symmetry to be based on $U(1)$ and $\mathbb{Z}_{2}$ subfactors. Even though there are $\mathbb{Z}_{n}$ subfactors in the family symmetry group, the mechanism is still identical to that of FN and therefore henceforth we will still refer to this as a FN mechanism. The $U(1)$ and $\mathbb{Z}_{2}$ subfactors may be a consequence of multiple spontaneous symmetry breaking from a larger GUT group G, valid at even higher energy scales than the $SU(5)$ scale. In this paper we do not address the actual generation of the family symmetry group from a higher GUT, since our main aim is to capture features in the quark and lepton masses and mixing angles. Thus, one major motivation for considering the Abelian groups as candidates for family symmetries is their ubiquitousness in any theory that has multiple symmetry breakings. 
\par
Now in the Lagrangian of the low energy effective theory, to write Yukawa coupling terms invariant under the family symmetry we must take the fermion fields to be charged under the family symmetry group. The mass terms in the Lagrangian are a result of non-renormalizable terms of the form
\be
\mathcal{L}_{~Y}=
\left(\frac{\eta_{a}}{\Lambda}\right)^{p_{ij}}\left(\frac{\eta_{b}}{\Lambda}\right)^{q_{ij}}\ldots~\left(\frac{\eta_{c}}{\Lambda}\right)^{r_{ij}}~\psi_{i}~\cdot~\chi_{j}~H_{(R)}
\label{fnnrt}
\eq
where $\Lambda$ is some characteristic energy scale at which the Yukawa coupling constants are generated,  $\psi$ and $\chi$ are the fermion representations in the model and $H_{(R)}$ is the Higgs scalar in representation R.  The subscripts $(i,~j,~..)$ refer to the generation indices and the indices $(a,~b,~...)$ refer to the scalar fields (flavons), one for each subfactor in the family symmetry group. Terms like these can be viewed as part of an effective Lagrangian that results after integrating out heavy fermion fields $\Psi_i$ with masses of the order of $\Lambda$ as seen in Fig.~(\ref{fn_diag}). Without loss of generality each $\eta$ flavon field is charged $-1$ under the respective family group subfactor. The Higgs scalars are assumed to carry no family symmetry charges so that the Higgs potential is unaffected. The invariance of the Yukawa term under the family symmetry implies
\be
\begin{split}
p_{ij}~&=~x_{i}\oplus x^{'}_{j}~~~,\\
q_{ij}~&=~y_{i}\oplus y^{'}_{j}~~~,\\
&~~\vdots\\
r_{ij}~&=~z_{i}\oplus z^{'}_{j}~~~,
\end{split}
\eq
where $(x_{i},~ x^{'}_{j};~ y_{i},~ y^{'}_{j};~\ldots)$ are the charges of the fermion representations under each subfactor of the family symmetry group. The symbol $\oplus$ denotes either ordinary addition if the subfactor is $U(1)$ or modulo addition for cyclic group subfactors. With this understanding, from the above expression, it is readily seen that the $U(1)$ charges satisfy the sum rule 
\be
p_{ij}+p_{ji}~=~p_{ii}+p_{jj}~~~,
\label{usumrule}
\eq
while the $\mathbb{Z}_{n}$ charges satisfy the sum rule
\be
q_{ij}+q_{ji}~=~q_{ii}+q_{jj}~({\rm mod}~n)~~~.
\label{zsumrule}
\eq
Thus we note that the $\mathbb{Z}_{n}$ subfactors may be used to selectively suppress the diagonal or off-diagonal elements in a $2\times2$ matrix block. This property will prove useful to us when we impose hierarchy in the quark sector.
 
 \begin{figure}
\begin{center}
\includegraphics[width=8.00cm,angle=0]{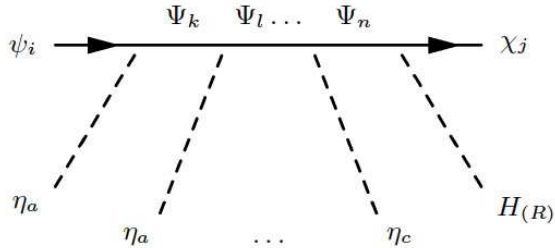}
\end{center}
\caption{Froggatt-Nielsen diagram that leads to effective Yukawa terms as in Eq.\ (\ref{fnnrt})}
\label{fn_diag}
\end{figure}

  The Yukawa coupling constants are generated when the flavon scalar fields,  $\eta$, get their VEVs. From Eq.\ (\ref{fnnrt}) we note that the actual value of the Yukawa constants depend on the fermion charges and the value of the VEVs. We do not try to incorporate  CP-violating phases in any of our analyses. Another point to note is that if the theory is assumed to be supersymmetric, holomorphy of the superpotential stipulates that $p_{ij},~q_{ij}\ldots~r_{ij}~\geqslant~0$ and be integers \cite{fnm}. For the model that we construct we will require that the sum of the fermion family charges be positive integers to be consistent with holomorphy. This is a major theoretical constraint when we pick charges for the various representations of $SU(5)$. Also in models where the horizontal symmetry is gauged, there may be anomalies present in the model \cite{fnm}. The anomalies may be made amenable to cancellation by requiring that the family symmetry commutes with the $SU(5)$ generators. Hence we must assign the \textit{same} charge under the family symmetry to all the fermion fields in a multiplet of $SU(5)$ (for a given generation). Thus the Froggatt-Nielsen mechanism can be used to impose further relations between Yukawa terms, including those of different generations. In this way after electroweak symmetry breaking one can then obtain logarithmically distributed fermion masses with the $U(1)_{F}$ and $\mathbb{Z}_{2}$ charges being ${\cal O}(1)$ parameters--that is, without much fine tuning. One of the main challenges of using this mechanism to build a model of flavor is choosing the right charges without much fine tuning so that one not only gets the fermion mass hierarchy correct,
but also the mixings after moving to the mass eigenstate basis.
 \end{section}


\begin{section}{An SU(5) Toy Model with Abelian Family Symmetry}
 \par
 We will now consider as an example an $SU(5)$ GUT model along with a $U(1)_{F}$ family symmetry. The hierarchical structure in the quark and charged lepton sectors is implemented using three $\mathbb{Z}_{2}$ symmetries that suppress the off-diagonal elements compared to the diagonal elements. This suppression of the off-diagonal elements is a \textit{sufficient} condition for the mixing angles to be small in the quark sector and hence also in the charged lepton sector. Let us label these discrete groups as $\mathbb{Z}_{2}$, $\mathbb{Z}^{'}_{2}$ and $\mathbb{Z}^{'''}_{2}$. It is possible that this suppression of the off-diagonal elements is a consequence of some other mechanism, but in this model we try to fold that ignorance into the  $\mathbb{Z}_{2}$  charges. The toy model is also constructed to be consistent with superpotential holomorphy, coming from supersymmetry considerations  \cite{fnm}.
 \par
In the quark sector it is sufficient to consider the hierarchies within each charge sector separately. The contributions to the up quark masses come from the $\bm{ 10\otimes10}$ fermion representations while those that contribute to the down quark masses come from the $\bm{ 10~\otimes5^{*}}$ fermion representations. Up to $\mathcal{O}(1)$ factors the Yukawa terms thus are
\be
\begin{split}
\mathcal{L}^{d}_{~Y}&\thicksim
\left(\frac{\eta_{a}}{\Lambda}\right)^{p_{ij}}\left(\frac{\eta_{b}}{\Lambda}\right)^{q_{ij}}\left(\frac{\eta_{c}}{\Lambda}\right)^{r_{ij}}\left(\frac{\eta_{d}}{\Lambda}\right)^{s_{ij}}Q_{i}~\bar{d}_{j}~H~~~,\\
\mathcal{L}^{u}_{~Y}&\thicksim\left(\frac{\eta_{a}}{\Lambda}\right)^{l_{ij}}\left(\frac{\eta_{b}}{\Lambda}\right)^{m_{ij}}\left(\frac{\eta_{c}}{\Lambda}\right)^{n_{ij}}\left(\frac{\eta_{d}}{\Lambda}\right)^{o_{ij}}Q_{i}~\bar{u}_{j}~ \tilde{H}~~~,
\end{split}
\label{quarkyukawas}
\eq
with $Q_{L}~\epsilon~\bm{(3,2)}~,~\bar{u}_{R}~\epsilon~\bm{ (3^{*}, 1)}$ in the $\bm{ 10}$ and $\bar{d}_{R}~\epsilon~\bm{(3^{*},1)}$ assigned to the $\bm{ 5^{*}}$. $H~\epsilon~\bm{ (1, 2)}$ is the usual SM Higgs doublet. $\tilde{H}$ is defined as $\tilde{H}=i\sigma_{2}~H^{*}$. The subscripts $a,~b,~c$ and $d$ on the flavon fields signifies their association to either the $U(1)_{F}$ or $\mathbb{Z}_{2}$ subfactors. Likewise $(p_{ij},~q_{ij},~r_{ij},~s_{ij})$ and $(l_{ij},~m_{ij},~n_{ij},~o_{ij})$ denote the sum of the $U(1)_{F}$ or $\mathbb{Z}_{2}$ charges respectively of the fermions in the Yukawa term. The labels $i,~j$ denote the generations. 
\par
Now we \textit{assume}  that the quark sector is `hierarchical' in the sense that the matrices that diagonalize the up and down quark matrices are themselves very close to the identity matrix. This means that the matrix may to zeroth approximation be represented as
\be
\mathcal{Y}_{U}~\thicksim~
m_{u}
\begin{pmatrix}
1\\
\emptyset\\
\emptyset
\end{pmatrix}
\cdot
\begin{pmatrix}
1&\emptyset&\emptyset
\end{pmatrix}
~\oplus~~m_{c}
\begin{pmatrix}
\emptyset\\
1\\
\emptyset
\end{pmatrix}
\cdot
\begin{pmatrix}
\emptyset&1&\emptyset
\end{pmatrix}
~\oplus~~
m_{t}
\begin{pmatrix}
\emptyset\\
\emptyset\\
1
\end{pmatrix}
\cdot
\begin{pmatrix}
\emptyset& \emptyset&1
\end{pmatrix}~~~,
\eq
where $\emptyset$ denotes an entry that is relatively suppressed due to a large power of the Cabibbo angle. Then from the up quark mass ratios in Eq.\ (\ref{qrel}) up to $\mathcal{O}(1)$ factors we have the $+~\frac{2}{3}$ quark Yukawa matrix
\be
\mathcal{Y}_{U}~\thicksim~
\begin{pmatrix}
\lambda^{8} &\ldots&\ldots \\
\ldots& \lambda^{4}&\ldots \\
\ldots&\ldots& 1
\end{pmatrix}~~~.
\label{yuplcal}
\eq
Since the above Yukawa matrix comes from the $\bm{ 10_{i}\otimes10_{j}}$ term it fixes the $U(1)_{F}$ charge of the $\bm{ 10}$ under the three generations as
\be
\bm{10}~~~:~~~~\text{I}~~(4,1^{'})~~~;~~~~\text{II}~~(2,1^{''})~~~;~~~~\text {III}~~(0,1^{'''})~~~.
\label{tencharg}
\eq
The three  $\mathbb{Z}_{2}$ charges are fixed by the requirement that the off-diagonal elements must be suppressed relative to the diagonal elements. The requirement of holomorphy is clearly satisfied for the above choices of the charges.
\par
The down quark Yukawa matrix comes from terms of the form $\bm{ 10_{i}~\otimes5^{*}_{j}}$. Unlike the up quark matrix, the down quark matrix in general is not symmetric or even \textit{normal} (see, for example, \cite{mw}). From the matrices $\mathcal{Y}_{D}\mathcal{Y}_{D}^T$ and $\mathcal{Y}_{D}^{T}\mathcal{Y}_{D}$, the assumption of hierarchy implies that the $-\frac{1}{3}$ quark Yukawa matrix decomposition must be again of the form

\be
\mathcal{Y}_{D}~\thicksim~
m_{d}
\begin{pmatrix}
1\\
\emptyset\\
\emptyset
\end{pmatrix}
\cdot
\begin{pmatrix}
1&\emptyset&\emptyset
\end{pmatrix}
~\oplus~~m_{s}
\begin{pmatrix}
\emptyset\\
1\\
\emptyset
\end{pmatrix}
\cdot
\begin{pmatrix}
\emptyset&1&\emptyset
\end{pmatrix}
~\oplus~~
m_{b}
\begin{pmatrix}
\emptyset\\
\emptyset\\
1
\end{pmatrix}
\cdot
\begin{pmatrix}
\emptyset& \emptyset&1
\end{pmatrix}~~~.
\eq

Now based on the down quark ratios and the estimate of the $m_{b}/m_{t}$ ratio at the GUT scale we may write the down quark Yukawa matrix as

\be
\mathcal{Y}_{D}~\thicksim~
\begin{pmatrix}
\lambda^{7} &\ldots&\ldots \\
\ldots&\lambda^{5}&\ldots \\
\ldots&\ldots& \lambda^{3}
\end{pmatrix}~~~
\label{ydownlcal}
\eq
where again there might be $\mathcal{O}(1)$ factors preceding each term. The $U(1)_{F}$ and  $\mathbb{Z}_{2}$  charges are again set by the mass ratios and `hierarchy' assumption to be
\be
\bm{ 5^{*}}~~~~:~~~~~\text{I}~~(3, 1^{'})~~~;~~~~\text{II}~~(3, 1^{''})~~~;~~~~\text{III}~~(3,1^{'''})~~~.
\label{fivescharg}
\eq
Thus we obtain equal $U(1)_{F}$ charges for $\bm{ 5^{*}}$ across the three generations. It is to be noted that this is a consequence of the fact that the logarithmic mass separation in the down quark sector is approximately half that in the up quark sector as we had assumed in Eq.\ (\ref{qrel}). As we shall see, this will be an important result when constructing the light neutrino Yukawa matrix. Again the  $\bm{ 5^{*}}$ charges are consistent with the holomorphy of the superpotential.
\par
We now turn our attention to the family symmetries and the minimal number of flavon scalar fields that are required in the model apart from the Higgs sector scalars. There must be one scalar field corresponding to each of the family symmetry subfactor groups and they attain a VEV at some characteristic energy scale that breaks the respective horizontal symmetry subfactor. Let us label these scalar fields and their charges under $U(1)_{F}\otimes\mathbb{Z}{'}_{2}\otimes\mathbb{Z}^{''}_{2}\otimes\mathbb{Z}^{'''}_{2}$ as
\be
\eta:~(-1)~~;~~~
\eta^{'}:~(-1^{'})~~;~~~
\eta^{''}:~(-1^{''})~~;~~~
\eta^{'''}:~(-1^{'''})~~~.
\eq 
Again we parametrize the characteristic energy scales at which these fields attain their VEVs by the Cabibbo angle. In this model we take for the $U(1)_{F}$ flavon field
\be
\left \langle \frac{\eta }{\Lambda}\right\rangle_{0}\thicksim~\lambda~~~,
\eq
and for the three $\mathbb{Z}_{2}$ flavons let
\be
 \left \langle \frac{\eta^{'} }{\Lambda}\right\rangle_{0}\thicksim ~\lambda^{\alpha}~~;~~~
\left \langle \frac{\eta^{''} }{\Lambda}\right\rangle_{0}\thicksim ~\lambda^{\beta}~~;~~~
\left \langle \frac{\eta^{'''} }{\Lambda}\right\rangle_{0}\thicksim~\lambda^{\gamma}~~~.
 \label{ztvev}
\eq
The values of these VEVs must be chosen so as to make the quark sector `hierarchical' with suppressed off-diagonal entries. Here $\Lambda$ is some characteristic energy scale associated with the breaking of the horizontal symmetry leading to the Yukawa coupling constants at lower energies. The three  $\mathbb{Z}_{2}$ flavon VEVs have been taken to be different in the above, since \textit{a priori} there is no reason to assume them to be the same or related in any way. We will comment further on the values of the  $\mathbb{Z}_{2}$ flavon VEVs when we discuss the constraints in the charged lepton and neutrino sectors. For now we just assume that the values of the VEVs parametrized by $\alpha,~\beta,~\gamma$ are sufficient to enforce a suppression of the off-diagonal elements in the quark Yukawa matrices. The Yukawa matrices in terms of these parameters are

\be
\begin{split}
&\mathcal{Y}_{U}~\thicksim~
\begin{pmatrix}
\lambda^{8} &\lambda^{6+\alpha+\beta}&\lambda^{4+\alpha+\gamma} \\
\lambda^{6+\alpha+\beta}& \lambda^{4}&\lambda^{2+\gamma+\beta} \\
\lambda^{4+\alpha+\gamma}&\lambda^{2+\gamma+\beta}& 1
\end{pmatrix}~~~,\\
~ \\
&\mathcal{Y}_{D}~\thicksim~
\begin{pmatrix}
\lambda^{7} &\lambda^{7+\alpha+\beta}&\lambda^{7+\alpha+\gamma} \\
\lambda^{5+\alpha+\beta}&\lambda^{5}&\lambda^{5+\gamma+\beta} \\
\lambda^{3+\alpha+\gamma}&\lambda^{3+\gamma+\beta}& \lambda^{3}
\end{pmatrix}~~~.
\label{yquarkvev}
\end{split}
\eq

\par
The CKM matrix is constructed from the `left' matrices that diagonalize $\mathcal{Y}_{D}$ and $\mathcal{Y}_{U}$. Bidiagonalizing the matrices in  Eqs.\ (\ref{yuplcal}) and (\ref{ydownlcal}) gives

\be
\begin{split}
&\text{Diag}~(m_{u},~m_{c},~m_{t})~\simeq~\mathcal{U}^{U}_{L}~\mathcal{Y}_{U}~\mathcal{U}^{U\, \dag}_{R}~~~,\\
&\text{Diag}~(m_{d},~m_{s},~m_{b})~\simeq~\mathcal{U}^{D}_{L}~\mathcal{Y}_{D}~\mathcal{U}^{D\, \dag}_{R}~~~.
\end{split}
\eq
In the above, for small mixing angles, the diagonalizing matrices \cite{lhar} to lowest order may be expressed in terms of the VEV parameters $\alpha,~\beta$ and $\gamma$ as
\be
\begin{split}
&\Bigl\lvert\mathcal{U}^{U}_{L}\Bigr\rvert~\thicksim~
\begin{pmatrix}
1 &\lambda^{2+\alpha+\beta}& \lambda^{4+\alpha+\gamma} \\
\lambda^{2+\alpha+\beta}&1& \lambda^{2+\beta+\gamma}\\
\lambda^{2+\alpha+\beta}& \lambda^{2+\beta+\gamma}& 1
\end{pmatrix}~~~,\\
&\Bigl\lvert\mathcal{U}^{D}_{L}\Bigr\rvert~\thicksim~
\begin{pmatrix}
1 &\lambda^{2+\alpha+\beta}+\lambda^{2+\alpha+\gamma}& \lambda^{4+\alpha+\gamma} \\
\lambda^{2+\alpha+\beta}+\lambda^{2+\alpha+\gamma}&1& \lambda^{2+\beta+\gamma}\\
\lambda^{2+\alpha+\beta}& \lambda^{2+\beta+\gamma}& 1
\end{pmatrix}~~~.
\end{split}
\label{luddiagvev}
\eq
The textures of both the up and down quark diagonalizing matrices come out to be the same up to undetermined $\mathcal{O}(1)$ factors. Since by assumption the VEV parameters are chosen so as to impose hierarchy the above matrices may be equivalently viewed as having the textures
\be
\begin{split}
&\Bigl\lvert\mathcal{U}^{U}_{L}\Bigr\rvert~\simeq~
\begin{pmatrix}
1 &\emptyset& \emptyset \\
\emptyset&1& \emptyset\\
\emptyset & \emptyset& 1
\end{pmatrix}~~~,\\
&\Bigl\lvert\mathcal{U}^{D}_{L}\Bigr\rvert~\simeq~
\begin{pmatrix}
1 & \emptyset& \emptyset \\
\emptyset&1& \emptyset \\
\emptyset& \emptyset& 1
\end{pmatrix}~~~.
\end{split}
\label{ldiag00}
\eq
 In the above, $\emptyset$ is used to denote a term that is suppressed with respect to the diagonal entry since it has $\lambda$ raised to a large power. As mentioned previously, we will explicitly give a bound on the possible values of the parameters $\alpha,~\beta,~\gamma$ when we discuss the neutrino sector. At this stage they are arbitrary, but sufficiently large to enforce hierarchy. Note that we do not use the idea of holomorphic zeroes \cite{fnm} in the Yukawa matrices to suppress entries and generate texture zeroes. From Eq.\ (\ref{ldiag00}) we may construct the CKM matrix as

\be
\Bigl\lvert\mathcal{U}_{CKM}\Bigr\rvert \simeq \Bigl\lvert\mathcal{U}^{U}_{L}\Bigr\rvert~\Bigl\lvert\mathcal{U}^{D\, \dag}_{L}\Bigr\rvert\simeq
\begin{pmatrix}
1 & \emptyset& \emptyset\\
\emptyset&1& \emptyset \\
\emptyset& \emptyset& 1
\end{pmatrix}~~~.
\label{ckmlcal}
\eq
It must again be commented that in this toy model the finer details of the CKM matrix as embodied in the Wolfenstein parametrization \cite{wolfen} and CP phases are not obtained. But using the $U(1)_{F}$ and $\mathbb{Z}_{2}$ horizontal symmetries we have constructed to first approximation the CKM matrix as an identity matrix. 
\par
In the quark sector the qualitative features of the quark mass hierarchy and CKM matrix are obtained.  By the imposition of hierarchy through discrete symmetries, the up quark masses are in the ratio
\be
m_{u}~:~m_{c}:m_{t}~\simeq~\mathcal{O}(\lambda^{8})~:~\mathcal{O}(\lambda^{4})~:~\mathcal{O}(1)~~~.
\eq
Similarly the down quark mass ratio is
\be
m_{d}~:~m_{s}:m_{b}~\simeq~\mathcal{O}(\lambda^{4})~:~\mathcal{O}(\lambda^{2})~:~\mathcal{O}(1)~~~,
\eq
with
\be
\frac{m_{b}}{m_{t}}~\simeq~\mathcal{O}(\lambda^{3})~~~,
\eq
and the CKM matrix is

\be
\Bigl\lvert\mathcal{U}_{CKM}\Bigr\rvert ~\simeq~
 \mathbb{I}_{3 \times 3}~~~.
\eq

\par
In the charged lepton sector the Yukawa terms are of the form $\bm{5^{*}_{i}\otimes 10_{j}}$. This is a consequence of $l_{L}~\epsilon~\bm{ (1, 2)}$ belonging to $\bm{5^{*}}$ and $\bar{e}_{R}~\epsilon~\bm{(1,1)}$ to the $\bm{10}$ of $SU(5)$.  The Yukawa terms in the usual notation are of the form
\be
\mathcal{L}^{l^{\pm}}_{~Y}\thicksim
\left(\frac{\eta_{a}}{\Lambda}\right)^{w_{ij}}\left(\frac{\eta_{b}}{\Lambda}\right)^{x_{ij}}\left(\frac{\eta_{c}}{\Lambda}\right)^{y_{ij}}\left(\frac{\eta_{d}}{\Lambda}\right)^{z_{ij}}~l_{i}~\bar{e}_{j} ~H~~~,
\eq
up to possible $\mathcal{O}(1)$ factors. The Yukawa constant matrix coming from the $\bm{5^{*}_{i}\otimes 10_{j}}$ term  is just the transpose of the down quark Yukawa matrix :
\be
\mathcal{Y}_{l^{\pm}}~\thicksim~
\mathcal{Y}^{T}_{D}~\thicksim~
\begin{pmatrix}
\lambda^{7} &\lambda^{7+\alpha+\beta}&\lambda^{7+\alpha+\gamma} \\
\lambda^{5+\alpha+\beta}&\lambda^{5}&\lambda^{5+\gamma+\beta} \\
\lambda^{3+\alpha+\gamma}&\lambda^{3+\gamma+\beta}& \lambda^{3}
\end{pmatrix}^{T}~~~.
\label{yleptonlcal}
\eq

Since we have constructed the quark sector to be `hierarchical' and since the charged lepton Yukawa matrix in $SU(5)$ is just the transpose of the down quark matrix, the charged lepton Yukawa matrix is also `hierarchical' :
\be
\mathcal{Y}_{l^{\pm}}~\thicksim~
m_{e}
\begin{pmatrix}
1\\
\emptyset\\
\emptyset
\end{pmatrix}
\cdot
\begin{pmatrix}
1&\emptyset&\emptyset
\end{pmatrix}
~\oplus~~m_{\mu}
\begin{pmatrix}
\emptyset\\
1\\
\emptyset
\end{pmatrix}
\cdot
\begin{pmatrix}
\emptyset&1&\emptyset
\end{pmatrix}
~\oplus~~
m_{\tau}
\begin{pmatrix}
\emptyset\\
\emptyset\\
1
\end{pmatrix}
\cdot
\begin{pmatrix}
\emptyset& \emptyset&1
\end{pmatrix}~~~.
\eq

One point to note is that the left diagonalizing matrix of the charged leptons is the diagonalizing matrix that appears on the right of the down quark Yukawa matrix. This implies that the off-diagonal suppression in the down Yukawa matrix must be sufficient to make the right diagonalizing matrix also close to a unit matrix. As mentioned before when we discuss the neutrino sector we will put bounds on the possible values of the VEV parameters. Thus the contribution to the PMNS lepton mixing matrix from the charged lepton sector will be a left diagonalizing matrix that is very close to unity. It is also to be commented that in $SU(5)$, as is well known, the mass hierarchy in the charged leptons closely follows the mass hierarchy in the down quarks. Poor mass relations such as
\be
\begin{split}
m_{\mu}&\simeq~m_{s}~~~,\\
 m_{e}&\simeq~m_{d}~~~,
\end{split}
\label{su5relations}
\eq
are a generic problem in minimal $SU(5)$ GUTs and may be improved by extending the Higgs sector which will contribute $\mathcal{O}(1)$ factors as Clebsch-Gordan coefficients. A classic example of this is the $SU(5)$ model of Georgi and Jarlskog \cite{geja} which extends the `minimal' $SU(5)$ model with the Higgs in the $\bm{5}$ representation with a Higgs in the $\bm{45}$ representation. This leads to the improved relations
\be
\begin{split}
m_{\mu}&\simeq~3~m_{s}~~~,\\
 m_{e}&\simeq~\frac{m_{d}}{3}~~~.
\end{split}
\label{imp_relations}
\eq
\par
Thus in the model that we pursue the charged lepton mass ratios come out naturally as
\be
m_{e}:m_{\mu}:m_{\tau}~\simeq~\mathcal{O}(\lambda^{4}):\mathcal{O}(\lambda^{2}):\mathcal{O}(1)~~~,
\eq
due to the imposition of hierarchy in the down quark sector. Extending the Higgs sector is likely to improve these ratios. Also due to the above-mentioned property of SU(5) the contribution to the lepton mixing matrix from the charged leptons, again assuming small mixing angles, is
\be
\Bigl\lvert\mathcal{U}^{l\pm}_{L}\Bigr\rvert~\thicksim~
\begin{pmatrix}
1 &\lambda^{\alpha+\beta}& \lambda^{\alpha+\gamma} \\
\lambda^{\alpha+\beta}&1& \lambda^{\beta+\gamma}\\
\lambda^{\alpha+\gamma}& \lambda^{\beta+\gamma}& 1
\end{pmatrix}\simeq
\begin{pmatrix}
1 &\emptyset& \emptyset \\
\emptyset&1& \emptyset \\
\emptyset& \emptyset& 1
\end{pmatrix}~~~,
\label{ulepL}
\eq
where again $\emptyset$ denotes suppressed entries.
\par
Turning now to the neutrino sector we accommodate three right-handed neutrinos ($N_{R}$) in the $\bm{ 1}$ representation of $SU(5)$. In our study it is assumed that the smallness of neutrino masses is explained by the Type-I see-saw mechanism \cite{see-saw}. In this context the Yukawa terms in the Lagrangian that contribute to the light neutrino masses may be written as Dirac and Majorana terms \cite{see-saw}

\be
\begin{split}
\mathcal{L}^{D}_{Y}&\thicksim\left(\frac{\eta_{a}}{\Lambda}\right)^{e_{ij}}\left(\frac{\eta_{b}}{\Lambda}\right)^{f_{ij}}\left(\frac{\eta_{c}}{\Lambda}\right)^{g_{ij}}\left(\frac{\eta_{d}}{\Lambda}\right)^{h_{ij}}~l_{i}\bar{N}^{R}_{j}\tilde{H}~~~,\\
\mathcal{L}^{N}_{Y}&\thicksim\left(\frac{\eta_{a}}{\Lambda}\right)^{t_{ij}}\left(\frac{\eta_{b}}{\Lambda}\right)^{u_{ij}}\left(\frac{\eta_{c}}{\Lambda}\right)^{v_{ij}}\left(\frac{\eta_{d}}{\Lambda}\right)^{k_{ij}}~\frac{\Lambda_R}{2}N^{R}_{i}N^{R}_{j}~~~,
\end{split}
\label{seeswlag}
\eq
up to $\mathcal{O}(1)$ factors from the non-renormalizable terms. $\Lambda_{R}$ is the characteristic energy scale of the right-handed Majorana neutrinos or equivalently the see-saw scale. In the Type-I see-saw the light neutrino masses are generated by the expression
\be
\mathcal{Y}_{\nu}~\simeq~-\mathcal{Y}^{D}_{\nu}(\mathcal{Y}^{R}_{\nu})^{-1}(\mathcal{Y}^{D}_{\nu})^{T}~~~,
\label{seesw}
\eq
where $\mathcal{Y}_{\nu}$ is the light neutrino Yukawa matrix, $\mathcal{Y}^{D}_{\nu}$ is the Dirac neutrino Yukawa matrix and $\mathcal{Y}^{R}_{\nu}$ is the right-handed Majorana neutrino Yukawa matrix. Now, in the basis where the charged lepton Yukawa matrix is diagonal the light neutrino Yukawa matrix texture
 \be
\mathcal{Y}_{\nu}~\thickapprox~~~
\begin{pmatrix}
a &b& b \\
b&b+c&a-c \\
b &a-c& b+c
\end{pmatrix}~~~,
\label{nutex}
\eq
leads immediately to the PMNS matrix in Eq.\ (\ref{pmns}). This may be observed from the decomposition
\be
\mathcal{Y}_{\nu}~\thicksim~
(a-b)
\begin{pmatrix}
\frac{-2}{\sqrt{6}}\\
~~~\frac{1}{\sqrt{6}}\\
~~~\frac{1}{\sqrt{6}}
\end{pmatrix}
\begin{pmatrix}
\frac{-2}{\sqrt{6}}&\frac{1}{\sqrt{6}}&\frac{1}{\sqrt{6}}
\end{pmatrix}
~\oplus~~
(a+2b)
\begin{pmatrix}
\frac{1}{\sqrt{3}}\\
\frac{1} {\sqrt{3}}\\
\frac{1} {\sqrt{3}}
\end{pmatrix}
\begin{pmatrix}
\frac{1} {\sqrt{3}}&\frac{1} {\sqrt{3}}&\frac{1} {\sqrt{3}}
\end{pmatrix}
~\oplus~~(2c+b-a)
\begin{pmatrix}
 ~0\\
\frac{-1}{\sqrt{2}}\\
\frac{1}{\sqrt{2}}
\end{pmatrix}
\begin{pmatrix}
0&\frac{-1}{\sqrt{2}}&\frac{1}{\sqrt{2}}
\end{pmatrix}~~.
\label{pmnsdecom}
\eq

It has been noted before that the above texture may be a consequence of a $\nu_{\mu}-\nu_{\tau}$ permutation symmetry  \cite{mutau}. We take Eqs.\ (\ref{numsq})\textendash(\ref{nuangles}) and Eq.\ (\ref{nutex}) as our guiding principles in constructing a phenomenologically viable neutrino sector in the toy model.

\par
Let us denote the $U(1)_{F}$ charges of the right-handed neutrinos ($N_{R}$) in the $\bm{ 1}$ across the three generations as 
\be
\bm{ 1}~:~~~~N^{R}_{1}~(e_1)~~;~~~N^{R}_{2} ~(e_2)~~;~~~N^{R}_{3}~(e_3)
\label{nrcharges}~~~.
\eq
The right-handed neutrinos are assumed to carry no $\mathbb{Z}_{2}$ charges. Note that in the $SU(5)$ model the Dirac Yukawa matrix $\mathcal{Y}^{D}_{\nu}$ in Eq.\ (\ref{seeswlag}) comes from the $\bm{ 5^{*}_{i}~\otimes1_{j}}$ terms while the Majorana Yukawa term  $\mathcal{Y}^{R}_{\nu}$ comes from the $\bm{ 1_{i}~\otimes1_{j}}$ terms.
If we denote the $U(1)_{F}$ charges of the $\bm{ 5^{*}}$ by $(x_1,~x_2,~x_3)$ then we have from Eqs.\ (\ref{ztvev}), (\ref{seesw}) and (\ref{nrcharges})
\be
\mathcal{Y}_{\nu}\thicksim \frac{v^{2}}{\Lambda_{R}}
\begin{pmatrix}
\lambda^{x_1+e_1+\alpha}&\lambda^{x_1+e_2+\alpha}&\lambda^{x_1+e_3+\alpha} \\
\lambda^{x_2+e_1+\beta}&\lambda^{x_2+e_2+\beta}&\lambda^{x_2+e_3+\beta} \\
\lambda^{x_3+e_1+\gamma}&\lambda^{x_3+e_2+\gamma} & \lambda^{x_3+e_3+\gamma}
\end{pmatrix}
\cdot
\left[\mathcal{Y}^{R}_{\nu}\right]^{-1}
\cdot
\begin{pmatrix}
\lambda^{x_1+e_1+\alpha}&\lambda^{x_1+e_2+\alpha}&\lambda^{x_1+e_3+\alpha} \\
\lambda^{x_2+e_1+\beta}&\lambda^{x_2+e_2+\beta}&\lambda^{x_2+e_3+\beta} \\
\lambda^{x_3+e_1+\gamma}&\lambda^{x_3+e_2+\gamma} & \lambda^{x_3+e_3+\gamma}
\end{pmatrix}^{T}
~~~,
\eq
where
\be
\left[\mathcal{Y}^{R}_{\nu}\right]^{-1}\thicksim
\frac{1}{\Lambda_{R}}
\begin{pmatrix}
\lambda^{2\,e_1}&\lambda^{e_1+e_2}&\lambda^{e_1+e_3} \\
\lambda^{e_2+e_1}&\lambda^{2\,e_2}&\lambda^{e_2+e_3} \\
\lambda^{e_3+e_1}&\lambda^{e_3+e_2} & \lambda^{2\,e_3}
\end{pmatrix}^{-1}
\thicksim
\frac{1}{\Lambda_{R}}
\begin{pmatrix}
\lambda^{-2\,e_1}&\lambda^{-(e_1+e_2)}&\lambda^{-(e_1+e_3)} \\
\lambda^{-(e_2+e_1)}&\lambda^{-2\,e_2}&\lambda^{-(e_2+e_3)} \\
\lambda^{-(e_3+e_1)}&\lambda^{-(e_3+e_2)} & \lambda^{-2\,e_3}
\end{pmatrix}~~~.
\eq
We are able to invert the seemingly singular matrix $\mathcal{Y}^R_\nu$ since there are suppressed $\mathcal{O}(1)$ factors in each term that render the matrix invertible. Using the above result we have
\be
\mathcal{Y}_{\nu}\thicksim \frac{v^{2}}{\Lambda_{R}}~
\begin{pmatrix}
\lambda^{2\,x_1+2\alpha}&\lambda^{x_1+x_2+\alpha+\beta}&\lambda^{x_1+x_3+\alpha+\gamma} \\
\lambda^{x_2+x_1+\beta+\alpha}&\lambda^{2\,x_2+2\beta}&\lambda^{x_2+x_3+\beta+\gamma} \\
\lambda^{x_3+x_1+\alpha+\gamma}&\lambda^{x_3+x_2+\gamma+\beta} & \lambda^{2\,x_3+2\gamma}
\end{pmatrix}~~~.
\eq
In the above expression we have powers like $2\alpha,~2\beta$ and $2\gamma$ because the see-saw scale is assumed to be lower than the scale $\Lambda$ where the flavons get a VEV and hence there is no $\mathbb{Z}_{2}$ addition in the see-saw formula. Thus we encounter the interesting property that in the $SU(5)$ model the light neutrino Yukawa matrix is generated by effective terms of the form
\be
\frac{v^{2}}{\Lambda_{R}}~(\bm{5}^{*}_{i}\otimes\bm{5}^{*}_{j})~~~.
\label{neueff}
\eq
Thus in the absence of $\mathbb{Z}_{2}$ charges for $\bm{1}$ and  holomorphic texture zeroes, the light neutrino Yukawa matrix is completely independent of the $\bm{1}$ charges. This may also be easily seen from the Feynman diagram of Fig.\ \ref{neu_mass} that leads to the light neutrino masses.

\begin{figure}
\begin{center}
\includegraphics[width=6.00cm,angle=0]{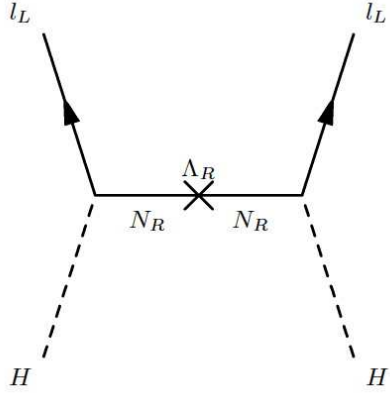}
\end{center}
\caption{Feynman diagram that generates light neutrino masses}
\label{neu_mass}
\end{figure}
\par
Using the above result along with the $\bm{ 5^{*}}$ charges in Eq.\ (\ref{fivescharg}) we get 

\be
\mathcal{Y}_{\nu}\thicksim \frac{v^{2}}{\Lambda_{R}}~\lambda^{6}~
\begin{pmatrix}
\lambda^{2\alpha}&\lambda^{\alpha+\beta}&\lambda^{\alpha+\gamma} \\
\lambda^{\beta+\alpha}&\lambda^{2\beta}&\lambda^{\beta+\gamma} \\
\lambda^{\alpha+\gamma}&\lambda^{\gamma+\beta} & \lambda^{2\gamma}
\end{pmatrix}~~~.
\label{gen_lightneu}
\eq
Now if we assume for example that $\alpha=\beta=\gamma$ then we get
\be
\mathcal{Y}_{\nu}~\thicksim~\frac{v^{2}}{\Lambda_{R}}~\lambda^{6+2\alpha}~
\begin{pmatrix}
1&1 & 1 \\
1&1&1 \\
1&1 & 1
\end{pmatrix}~~~.
\eq
This would correspond to the special case of $a=1$, $b=1$ and $c=0$ in Eq.\ (\ref{nutex}). Thus the light neutrino Yukawa matrix comes out naturally to be of a `democratic' structure as we had required. We note that this is strictly a consequence of the fact that the $\bm{ 5^{*}}$ charges across generations are all equal as seen from Eqs.\ (\ref{fivescharg}) and (\ref{neueff}). The equal  $\bm{ 5^{*}}$ charges are a result of the up quark sector's logarithmic mass spacing being twice that of the down quark sector. As an aside it must be pointed out that the earliest studies on a democratic Yukawa texture in the context of mass generation were by Nambu \cite{nambu} and Kaus and Meshkov \cite{kaumesh}. The above matrix has the $\mu-\tau$ symmetric texture that we require but it is immediately seen from Eq.\ (\ref{pmnsdecom}) that the above result is in conflict with Eq.\ (\ref{numsq}), since the mass hierarchies come out incorrectly. This requirement from neutrino oscillations and  $\mu-\tau$ symmetry motivates us to make the \textit{ansatz}
\be
\begin{split}
&\frac{\langle \eta^{'}\rangle }{\Lambda}~\thicksim~\lambda^{\alpha}~~~,\\
&\frac{\langle \eta^{''}\rangle }{\Lambda}~\thicksim~\frac{\langle \eta^{'''}\rangle }{\Lambda}~\thicksim~\lambda^{\beta}~~~,
\end{split}
\eq
which leads to the light neutrino texture
\be
\mathcal{Y}_{\nu}~\thicksim~\frac{v^{2}}{\Lambda_{R}}~\lambda^{6}~
\begin{pmatrix}
\lambda^{2\alpha}&\lambda^{\alpha+\beta}&\lambda^{\alpha+\beta} \\
\lambda^{\alpha+\beta}&\lambda^{2\beta}&\lambda^{2\beta} \\
\lambda^{\alpha+\beta}&\lambda^{2\beta} & \lambda^{2\beta}
\end{pmatrix}~~~.
\label{lightneu1}
\eq
It must be noted that choosing the VEVs of two of the $\mathbb{Z}_{2}$ subfactors to be different from the first clearly does not affect the result in Eq.\ (\ref{neueff}). It also allows us to properly account for the observed neutrino mass splittings. Comparing Eqs.\ (\ref{nutex}) and (\ref{lightneu1}) we may make, up to $\mathcal{O}(1)$ and common factors, the association
\be
\begin{split}
& a~\thicksim~\lambda^{6+2\alpha}~~~,\\
& b~\thicksim~\lambda^{6+\alpha+\beta}~~~,\\
& c~\thicksim~\lambda^{6+2\beta}~~~.
\end{split}
\eq
To make this association we have implicitly assumed that $\beta<\alpha$. We will comment on an alternative when we discuss the case of \textit{inverted} neutrino hierarchy. With this then we have for the light neutrino eigenvalues
\be
\begin{split}
&m_{\nu1}\thicksim\frac{v^{2}}{\Lambda_{R}}~~\lambda^{6}\left( \lambda^{2\alpha}-\lambda^{\alpha+\beta}\right)~~~,\\
&m_{\nu2}\thicksim\frac{v^{2}}{\Lambda_{R}}~~\lambda^{6}\left( \lambda^{2\alpha}+2\lambda^{\alpha+\beta}\right)~~~,\\
&m_{\nu3}\thicksim\frac{v^{2}}{\Lambda_{R}}~~\lambda^{6}\left( 2\lambda^{2\beta}-\lambda^{2\alpha}+\lambda^{\alpha+\beta}\right)~~~.
\end{split}
\label{neueig}
\eq
The neutrino mass squared differences in terms of the VEV parameters are
\be
\begin{split}
&\Delta_{\nu} m^{2}_{32}~\thicksim~\left(\frac{v^{2}}{\Lambda_{R}}~\lambda^{6}\right)^{2}~\left[ \left(2\lambda^{2\beta}-\lambda^{2\alpha}+\lambda^{\alpha+\beta}\right)^{2}~-~\left(2 \lambda^{\alpha+\beta}+\lambda^{2\alpha}\right)^{2}\right]~~~,\\
&\Delta_{\nu} m^{2}_{21}~\thicksim~\left(\frac{v^{2}}{\Lambda_{R}}~\lambda^{6}\right)^{2}~\left[ \left(\lambda^{2\alpha}+~2\lambda^{\alpha+\beta}\right)^{2}~-~\left( \lambda^{2\alpha}-\lambda^{\alpha+\beta}\right)^{2}\right]~~~.
\end{split}
\eq
Taking the ratio of the above mass squared differences and comparing with Eq.\ (\ref{numsq}) from neutrino oscillation data gives the condition
\be
\frac{4}{3}~ \lambda^{2(\beta-\alpha)}~\simeq~31~\Rightarrow~\alpha\simeq\beta+1~~~.
\label{alphaval}
\eq
We substitute back this bound into the quark sector results previously obtained. Assuming that the quark mixing angles are small (hierarchy assumption) we may explicitly derive \cite{lhar} up to $\mathcal{O}(1)$ factors  the diagonalizing matrices in terms of the VEV parameter $\beta$. We have the decomposition for the $+\frac{2}{3}$ charged quark sector
\be
\mathcal{Y}_{U}\thicksim
\lambda^{8}
\begin{pmatrix}
1\\
\lambda^{3+2\beta}\\
\lambda^{5+2\beta}
\end{pmatrix}
\begin{pmatrix}
1~\lambda^{3+2\beta}~\lambda^{5+2\beta}
\end{pmatrix}
\oplus \lambda^{4}
\begin{pmatrix}
-\lambda^{3+2\beta}\\
1\\
\lambda^{2+2\beta}
\end{pmatrix}
\begin{pmatrix}
-\lambda^{3+2\beta}~1~\lambda^{2+2\beta}
\end{pmatrix}
\oplus
1
\begin{pmatrix}
-\lambda^{5+2\beta}\\
-\lambda^{2+2\beta}\\
1
\end{pmatrix}
\begin{pmatrix}
-\lambda^{5+2\beta}~-\lambda^{2+2\beta}~1
\end{pmatrix}~,
\eq
which leads to the left diagonalizing matrix
\be
\Bigl\lvert\mathcal{U}^{U}_{L}\Bigr\rvert~\thicksim~
\begin{pmatrix}
1 &\lambda^{3+2\beta}& \lambda^{5+2\beta} \\
\lambda^{3+2\beta}&1& \lambda^{2+2\beta}\\
\lambda^{5+2\beta}&\lambda^{2+2\beta}& 1
\end{pmatrix}~~~.
\label{upmixvev}
\eq
Similarly,  considering the matrix $\mathcal{Y}_{D}.\mathcal{Y}_{D}^T$ we get the left diagonalizing matrix for the $-\frac{1}{3}$ charged quark sector, 
\be
\Bigl\lvert\mathcal{U}^{D}_{L}\Bigr\rvert~\thicksim~
\begin{pmatrix}
1 &\lambda^{3+2\beta}& \lambda^{5+2\beta} \\
\lambda^{3+2\beta}&1& \lambda^{2+2\beta}\\
\lambda^{5+2\beta}&\lambda^{2+2\beta}& 1
\end{pmatrix}~~~.
\label{downmixvev}
\eq
The two left diagonalizing matrices come out to be of the same texture for both the up and down quark sectors. Considering the matrix $\mathcal{Y}_{D}^{T}.\mathcal{Y}_{D}$ which is equivalent to $\mathcal{Y}_{l^{\pm}}.\mathcal{Y}_{l^{\pm}}^{T}$ gives
\be
\Bigl\lvert\mathcal{U}^{l^{\pm}}_{L}\Bigr\rvert~\thicksim~
\begin{pmatrix}
1 &\lambda^{1+2\beta}& \lambda^{1+2\beta} \\
\lambda^{1+2\beta}&1& \lambda^{2\beta}\\
\lambda^{1+2\beta}&\lambda^{2\beta}& 1
\end{pmatrix}~~~.
\label{chrglepmixvev}
\eq
The CKM matrix to lowest order comes out to be of the same texture as the up and down diagonalizing matrices
\be
\Bigl\lvert\mathcal{U}_{CKM}\Bigr\rvert~\thicksim~
\begin{pmatrix}
1 &\lambda^{3+2\beta}& \lambda^{5+2\beta} \\
\lambda^{3+2\beta}&1& \lambda^{2+2\beta}\\
\lambda^{5+2\beta}&\lambda^{2+2\beta}& 1
\end{pmatrix}~~~.
\label{1ckmmixvev}
\eq

We note that for any non-negative value for $\beta$ the $1-2$ CKM element is smaller than the $2-3$ CKM element. This texture is contrary to what is experimentally observed. We will comment further on this when we discuss an inverted neutrino mass hierarchy. To lowest order though the CKM matrix is an identity matrix. From Eqs.\ (\ref{upmixvev})-(\ref{chrglepmixvev}) and the requirement of `hierarchy' we get a bound on the VEV parameter (specifically from the charged lepton sector) 

\be
2\beta~\geq~1~~\Rightarrow~~\beta~\geq~\frac{1}{2}~~~.
\eq
We do not want the flavor symmetry breaking scale to be too low, since it would then affect low energy electroweak physics. This leads us to choose the lowest possible value for $\beta$, that is, the highest possible flavor symmetry breaking scale allowed, and to use a value of $\alpha$ consistent witn Eq.\ (\ref{alphaval}): 
\be
\begin{split}
&\alpha~=~\frac{3}{2}~~~,\\
&\beta~=~\frac{1}{2}~~~.
\label{vevchoice}
\end{split}
\eq
This leads to the light neutrino Yukawa matrix (up to $\mathcal{O}(1)$ factors)
\be
\mathcal{Y}_{\nu}~\thicksim~\frac{v^{2}}{\Lambda_{R}}~\lambda^{7}~
\begin{pmatrix}
\lambda^{2}& \lambda&\lambda \\
 \lambda&1&1 \\
 \lambda&1&1
\end{pmatrix}~~~,
\label{norm_hier_lightneu}
\eq
which is phenomenologically viable. In this model from Eq.\ (\ref{pmnsdecom}) the light neutrino mass ratio predictions are
\be
m_{\nu_{1}}~:~m_{\nu_{2}}~:~m_{\nu_{3}}~\simeq~\mathcal{O}(\lambda)~:~\mathcal{O}(\lambda)~:~\mathcal{O}(1)~~~.
\eq
Thus for the particular values of the flavon VEVs chosen, the toy model predicts that the neutrino mass spectrum is of the \textit{normal hierarchy} type. For the mass square differences
\be
\begin{split}
&\Delta_{\nu} m^{2}_{32}~\simeq~\mathcal{O}(10^{-3}~~{\text eV}^{2})~~~,\\
&\Delta_{\nu} m^{2}_{21}~\simeq~\mathcal{O}(10^{-5}~~{\text eV}^{2})~~~,
\end{split}
\label{numsq1}
\eq
with  $v\simeq174~{\text GeV}$ the model predicts
\be
\Lambda_{R}~\simeq~\mathcal{O}(10^{10}~~\text{GeV})~~~.
\eq

Although the see-saw scale comes out slightly lower than the GUT scale this is again quite consistent with experimental constraints and theoretical prejudices since the scale is not \textit{a priori} well defined (see, for example, \cite{{see-saw},{gouvea}}). The PMNS matrix without any phases or $\mathcal{O}(1)$ factors is to lowest order tri-bimaximal owing to the particular texture of the light neutrino matrix,
\be
\Bigl\lvert\mathcal{U}_{PMNS}\Bigr\rvert=\Bigl\lvert \mathcal{U}^{l^{\pm}\, \dag}_{L}~ \mathcal{U}^{\nu}_{L}\Bigr\rvert \simeq
\begin{pmatrix}
&\frac{2}{\sqrt{6}}+~\frac{\lambda^{2}}{\sqrt{6}} &~\frac{1}{\sqrt{3}} -~\frac{\lambda^{2}}{\sqrt{3}}&~\frac{\lambda^{2}}{\sqrt{2}}\\
&\frac{1}{\sqrt{6}}-\frac{\lambda}{\sqrt{6}}&~\frac{1}{\sqrt{3}}-\frac{\lambda}{\sqrt{3}}&~\frac{1}{\sqrt{2}}+\frac{\lambda}{\sqrt{2}} \\
&\frac{1}{\sqrt{6}}+\frac{\lambda}{\sqrt{6}} &~\frac{1}{\sqrt{3}}+\frac{\lambda}{\sqrt{3}}&~\frac{1}{\sqrt{2}}-\frac{\lambda}{\sqrt{2}}
\end{pmatrix}
\thicksim
\begin{pmatrix}
\frac{2}{\sqrt{6}} &\frac{1}{\sqrt{3}} & 0 \\
\frac{1}{\sqrt{6}} &\frac{1}{\sqrt{3}}&\frac{1}{\sqrt{2}} \\
\frac{1}{\sqrt{6}} &\frac{1}{\sqrt{3}}& \frac{1}{\sqrt{2}}
\end{pmatrix}~~~.
\label{pmns1}
\eq

\par
If instead of assuming that $\beta<\alpha$ we had assumed that $\beta>\alpha$ then we may again make the association up to common factors
\be
\begin{split}
& a~\thicksim~\lambda^{6+2\alpha}~~~,\\
& b~\thicksim~\lambda^{6+\alpha+\beta}~~~,\\
& b+c~\thicksim~a-c~\thicksim~\lambda^{6+2\beta}~~~.
\end{split}
\eq
 This association then leads to the light neutrino eigenvalues
\be
\begin{split}
&m_{\nu1}\thicksim\frac{v^{2}}{\Lambda_{R}}~\lambda^{6}\left( \lambda^{2\alpha}-\lambda^{\alpha+\beta}\right)~~~,\\
&m_{\nu2}\thicksim\frac{v^{2}}{\Lambda_{R}}~\lambda^{6}\left( \lambda^{2\alpha}+2\lambda^{\alpha+\beta}\right)~~~,\\
&m_{\nu3}\thicksim\frac{v^{2}}{\Lambda_{R}}~\lambda^{6}\left( \lambda^{2\beta}\right)~~~.
\end{split}
\label{case2neueig}
\eq
This choice clearly leads to an \textit{inverted} hierarchy in the neutrino sector since now $m_{\nu3}$ has a lesser value than $m_{\nu2}$ and $m_{\nu1}$. It must still be checked whether the neutrino oscillation data may be accommodated readily for reasonable values of the VEV parameters.  
The neutrino mass squared differences in terms of the VEV parameters are
\be
\begin{split}
&\Delta_{\nu} m^{2}_{32}~\thicksim~\left(\frac{v^{2}}{\Lambda_{R}}~\lambda^{6}\right)^{2}~\left[ ~\left(2 \lambda^{\alpha+\beta}+\lambda^{2\alpha}\right)^{2}-\left(\lambda^{2\beta}\right)^{2}\right]~~~,\\
&\Delta_{\nu} m^{2}_{21}~\thicksim~\left(\frac{v^{2}}{\Lambda_{R}}~\lambda^{6}\right)^{2}~\left[ \left(\lambda^{2\alpha}+~2\lambda^{\alpha+\beta}\right)^{2}~-~\left( \lambda^{2\alpha}-\lambda^{\alpha+\beta}\right)^{2}\right]~~~.
\end{split}
\eq
As we did for the previous case, taking the ratio of the above mass squared differences and comparing with Eq.\ (\ref{numsq}) gives
\be
\frac{1}{6}\left[ \lambda^{(\alpha-\beta)}+4\right]~\simeq~31~\Rightarrow~\beta\simeq\alpha+4~~~.
\label{case2alphaval}
\eq
This condition on the VEV parameters is again substituted back into the quark sector results. For small mixing angles \cite{lhar} this leads to the left diagonalizing matrices to lowest order for the up, down and charged lepton sectors :
\be
\begin{split}
&\Bigl\lvert\mathcal{U}^{U}_{L}\Bigr\rvert~\thicksim~
\begin{pmatrix}
1 &\lambda^{6+2\alpha}& \lambda^{8+2\alpha} \\
\lambda^{6+2\alpha}&1& \lambda^{10+2\alpha}\\
\lambda^{8+2\alpha}&\lambda^{10+2\alpha}& 1
\end{pmatrix}~~~,\\
&\Bigl\lvert\mathcal{U}^{D}_{L}\Bigr\rvert~\thicksim~
\begin{pmatrix}
1 &\lambda^{6+2\alpha}& \lambda^{8+2\alpha} \\
\lambda^{6+2\alpha}&1& \lambda^{10+2\alpha}\\
\lambda^{8+2\alpha}&\lambda^{10+2\alpha}& 1
\end{pmatrix}~~~,\\
&\Bigl\lvert\mathcal{U}^{l^{\pm}}_{L}\Bigr\rvert~\thicksim~
\begin{pmatrix}
1 &\lambda^{4+2\alpha}& \lambda^{4+2\alpha} \\
\lambda^{4+2\alpha}&1& \lambda^{8+2\alpha}\\
\lambda^{4+2\alpha}&\lambda^{8+2\alpha}& 1
\end{pmatrix}~~~.
\end{split}
\label{case2mixmatrices}
\eq
Because the hierarchical structure of the up and down quark mixing matrices the CKM matrix would again have the same texture as these matrices. That is,
\be
\Bigl\lvert\mathcal{U}_{CKM} \Bigr\rvert~\thicksim~
\begin{pmatrix}
1 &\lambda^{6+2\alpha}& \lambda^{8+2\alpha} \\
\lambda^{6+2\alpha}&1& \lambda^{10+2\alpha}\\
\lambda^{8+2\alpha}&\lambda^{10+2\alpha}& 1
\end{pmatrix}~~~.
\eq

 We note that in contrast to the case of normal neutrino hierarchy now the CKM matrix element in $1-2$ is greater than the $2-3$ element. So within the approximations of this model it seems that an inverted neutrino hierarchy gives a CKM matrix texture qualitatively closer to the observed CKM matrix. But again it is to be admitted that the $1-3$ element is wrong and the magnitudes of the CKM elements also come out incorrectly in the toy model. As before from the requirement of suppressed off-diagonal entries and small mixing angles we get a bound on the VEV parameter 
\be
4+2\alpha~\geq~1~~\Rightarrow~~\alpha~\geq~-\frac{3}{2}~~~.
\eq
Since the non-renormalizable terms are obtained after integrating out the heavy fermions ($\Psi$) we require that the VEV parameters ($\alpha,\beta$) be such that the flavon VEVs are smaller than the energy scale of the heavy fermions ($\Lambda$). This requires us to choose non-negative values for the VEV parameters. Choosing again the smallest value consistent with the bound gives
\be
\begin{split}
&\alpha~=~0~~~,\\
&\beta~=~4~~~.
\label{case2vevchoice}
\end{split}
\eq
With this choice the light neutrino Yukawa matrix is
\be
\mathcal{Y}_{\nu}~\thicksim~\frac{v^{2}}{\Lambda_{R}}~\lambda^{6}~
\begin{pmatrix}
1& \lambda^{4}&\lambda^{4} \\
 \lambda^{4}&\lambda^{8}& \lambda^{8} \\
 \lambda^{4}& \lambda^{8}& \lambda^{8}
\end{pmatrix}~~~,
\label{case2lightneu}
\eq
 So with the above choice of the VEV parameters from Eq.\ (\ref{case2neueig})
this toy model predicts the \textit{inverted} neutrino hierarchy
\be
m_{\nu_{1}}~:~m_{\nu_{2}}~:~m_{\nu_{3}}~\simeq~\mathcal{O}(1)~:~\mathcal{O}(1)~:~\mathcal{O}(\lambda^{8})~~~.
\eq
Now, from the neutrino mass squared differences and $v\simeq174~{\text GeV}$, the model predicts the see-saw scale to be
\be
\Lambda_{R}~\simeq~\mathcal{O}(10^{11}~~\text{GeV})~~~.
\eq
\par
The lepton mixing matrix is again tri-bimaximal. In this toy model the tri-bimaximal nature is strictly a consequence of the light neutrinos, with the charged lepton contribution being close to an identity matrix. This is a consequence of imposing a `hierarchy' in the quark sector which gets communicated to the charged lepton sector owing to the properties of the $SU(5)$ representations. The light neutrino Yukawa matrix is effectively generated from the equal $\bm{ 5^{*}}$ charges alone and this makes it naturally of a `democratic' type. Motivated by neutrino oscillation results we were able to pick the VEVs to readily incorporate the neutrino hierarchy and large mixing angles. The corrections to the strictly tri-bimaximal mixing in the toy model come from the charged lepton sector. Since the CKM matrix given by the toy model does not capture to next order the true texture in the $1-2$, $1-3$ and $2-3$ elements, the above result for the PMNS matrix must be considered only as an approximate first order prediction. 
 \end{section}
 

 \begin{section}{Problems in SU(5) GUTs with Abelian Family Symmetries}
 \par
 The toy model presented in the previous section captures to first approximation the mass hierarchy and mixing matrices in the quark and lepton sectors. When one attempts to include the finer details in the mass spectra and mixing angles some difficulties arise. To describe the fine structure without introducing large numbers of additional scalar fields, make further assumptions about the couplings, or otherwise drastically increasing the number of parameters has proved to be challenging. We briefly illustrate some of the difficulties by taking the toy model of the previous section and the Georgi-Jarlskog $SU(5)$ model \cite{geja} as examples.
 \par
 In the toy model that we explored, the choice of VEVs in Eqs.\ (\ref{vevchoice}) and (\ref{case2vevchoice})  led to the CKM mixing matrices
 \be
 \begin{split}
 &\Bigl\lvert\mathcal{U}^{(1)}_{CKM}\Bigr\rvert \thicksim
 \begin{pmatrix}
1& \lambda^{4}& \lambda^{6}~\\
\lambda^{4}&1& \lambda^{3}~\\
\lambda^{6}&\lambda^{3}& 1~
\end{pmatrix}~~~,\\
&\Bigl\lvert\mathcal{U}^{(2)}_{CKM}\Bigr\rvert \thicksim
 \begin{pmatrix}
1& \lambda^{6}& \lambda^{8}~\\
\lambda^{6}&1& \lambda^{10}~\\
\lambda^{8}&\lambda^{10}& 1~
\end{pmatrix}~~~.\\
\end{split}
\eq
The lepton mixing matrices in both cases are very close to tri-bimaximal
\be
\Bigl\lvert\mathcal{U}_{PMNS}\Bigr\rvert \thicksim
\ \begin{pmatrix}
&\frac{2}{\sqrt{6}}&\frac{1}{\sqrt{3}} &0\\
&\frac{1}{\sqrt{6}}&\frac{1}{\sqrt{3}}&\frac{1}{\sqrt{2}} \\
&\frac{1}{\sqrt{6}}&\frac{1}{\sqrt{3}}& \frac{1}{\sqrt{2}}
\end{pmatrix}
~~~.
\label{mixmcal}
\eq
Generically, in a minimal $SU(5)$ GUT with a family symmetry made up only of $U(1)$ subfactors, the PMNS matrix will come out to be unity.  This is because the left-diagonalizing matrices of the charged leptons and light neutrinos come out to be the same.  The $\mathbb{Z}_n$ subfactors that we used overcome this problem by imposing hierarchy on the charged leptons; they also set the neutrino mass hierarchy as noted previously.
\par
It is clear that the CKM matrix to next order does not agree well with that which is observed.  This is a generic problem in our model.  If we use the Yukawa matrices from Eq.\ (\ref{yquarkvev}) we obtain the CKM matrix for small mixing angles \cite{lhar}:
\be
\Bigl\lvert\mathcal{U}_{CKM}\Bigr\rvert \thicksim
\begin{pmatrix}
1 & \lambda^{2+\alpha+\beta} & \lambda^{4+\alpha+\gamma}\\
\lambda^{2+\alpha+\beta} & 1  & \lambda^{2+\beta+\gamma}\\
\lambda^{4+\alpha+\gamma} & \lambda^{2+\beta+\gamma} & 1
\end{pmatrix}~~~.
\label{ckm_gen}
\eq
Since $\alpha$, $\beta$, $\gamma > 0$ we see that $V_{us}$ comes out to be too small.  The requirement of $\mu$-$\tau$ symmetry fixes $\beta=\gamma$.  This sets $V_{ub}/V_{us}\thicksim \lambda^2$ which is in agreement with experiment.  However, folding in the neutrino masses causes problems.  Imposing a normal hierarchy sets $\alpha=\beta+1=\gamma+1$ giving $V_{cb}/V_{us}\thicksim \lambda^{-1}$ and $V_{ub}/V_{cb}\thicksim \lambda^3$.  In an inverted hierarchy $\alpha+4=\beta=\gamma$.  This gives $V_{cb}/V_{us}\thicksim \lambda^4$ and $V_{ub}/V_{cb}\thicksim \lambda^{-2}$.

This gives some insight into the difficulty of reconciling the neutrino masses and the CKM matrix within our model.  There are a few possible solutions. One may introduce one or more Higgs fields and charge them under the family symmetry.  This could serve to change the structure of the Yukawa matrices in Eq.\ (\ref{yquarkvev}) which lead directly to the CKM matrix of Eq.\ (\ref{ckm_gen}).  It may be that the $SU(5)$ GUT is not amenable to a description of the finer structure of the CKM and PMNS matrices.  Of course, enforcing that the quarks and charged leptons are `hierarchical' and that the light neutrinos are `democratic' could be wrong.

 \par
 We now explore the Georgi-Jarlskog model. As mentioned before, the Georgi-Jarlskog $SU(5)$ model incorporates improved quark-lepton relations
 \be
\begin{split}
m_{\tau}&\simeq~m_{b}~~~,\\
m_{\mu}&\simeq~3~m_{s}~~~,\\
m_{e}&\simeq~\frac{m_{d}}{3}~~~,
\end{split}
\label{gjrelations}
\eq
by introducing a  $\bm{45}$ Higgs representation. The Yukawa coupling terms are chosen to give the following textures for the quark and charged lepton matrices  \cite{geja}
\begin{align}
\mathcal{Y}_{U}~&=~
\begin{pmatrix}
0 &Y_{1'2'}& 0 \\
Y_{1'2'} &0& Y_{2'3'} \\
0 &Y_{2'3'}& Y_{3'3'}
\end{pmatrix}~~~,
\label{yup}
\\
\nonumber\\
\mathcal{Y}_{D}~&=~
\begin{pmatrix}
0 &Y_{2'1}& 0 \\
Y_{1'2} &Y_{2'2}&0 \\
0 &0& Y_{3'3}
\end{pmatrix}~~~,
\label{ydown}
\\
\nonumber\\
\mathcal{Y}_{l^{\pm}}~&=~
\begin{pmatrix}
0 &Y_{2'1}& 0 \\
Y_{1'2} &-3~Y_{2'2}&0 \\
0 &0& Y_{3'3}
\end{pmatrix}~~~.
\label{ylepton}
\end{align}

\par
Georgi and Jarlskog then note that for $Y_{1'2}=Y_{2'1}$,  $Y_{3'3}\gg Y_{2'2}\gg Y_{1'2}$  and $Y_{3'3'}\gg Y_{2'3'}\gg Y_{1'2'}$ in Eqs.\ (\ref{yup})--(\ref{ylepton}) the relations in Eq.\ (\ref{gjrelations}) are satisfied. Considering Eq.\ (\ref{qrel}) along with the mass eigenvalues of the above matrices \cite{geja}  we are immediately led to the identification
\be
\begin{split}
& Y_{1'2}\thicksim Y_{2'1}\thicksim Y_{1'2'}\thicksim\lambda^{7}~~~,\\
& Y_{3'3}\thicksim\lambda^{4}~,~ Y_{2'2}\thicksim\lambda^{6}~,~Y_{3'3'}\thicksim\lambda~,~ Y_{2'3'}\thicksim\lambda^{3}~~~,
\end{split}
\label{yassign}
\eq
with all the other Yukawa coupling constants zero. Now, using Eq.\ (\ref{yassign}) in Eqs.\ (\ref{yup})--(\ref{ylepton}) gives (neglecting any common factors and external $\mathcal{O}(1)$ factors from the non renormalizable FN terms)
\begin{align}
\mathcal{Y}_{U}~&\thicksim~
\begin{pmatrix}
0 &\lambda^{6}& 0 \\
\lambda^{6} &0& \lambda^{2} \\
0 &\lambda^{2}& 1
\end{pmatrix}~~~,
\label{yupl}
\\
\nonumber\\
\mathcal{Y}_{D}~&\thicksim~
\begin{pmatrix}
0 &\lambda^{6}& 0 \\
\lambda^{6} &\lambda^{5}&0 \\
0 &0& \lambda^{3}
\end{pmatrix}~~~,
\label{ydownl}
\\
\nonumber\\
\mathcal{Y}_{l^{\pm}}~&\thicksim~
\begin{pmatrix}
0 &\lambda^{6}& 0 \\
\lambda^{6} &-3~\lambda^{5}&0 \\
0 &0&\lambda^{3}
\end{pmatrix}~~~.
\label{yleptonl}
\end{align}

\par
The CKM matrix is constructed from the `left' matrices that diagonalize $\mathcal{Y}_{D}$ and $\mathcal{Y}_{U}$. Bidiagonalizing the matrices in  Eqs.\ (\ref{yupl}) and (\ref{ydownl}) gives (for $\lambda~\simeq~0.23$) 
\begin{align}
\Bigl\lvert\mathcal{U}^{U\, \dag}_{L}\Bigr\rvert~&\simeq~
\begin{pmatrix}
0.999 &0.04& 0.002 \\
0.04&0.998&0.04\\
0 &0.04& 0.999
\end{pmatrix}
~\thicksim~
\begin{pmatrix}
1 &0& 0 \\
0&1&0 \\
0 &0& 1
\end{pmatrix}~~~,
\label{ludiag}
\\
\nonumber\\
\Bigl\lvert\mathcal{U}^{D}_{L}\Bigr\rvert~&\simeq~
\begin{pmatrix}
0.982 &0.189& 0 \\
0.189&0.982&0\\
0 &0& 1
\end{pmatrix}
~\thicksim~
\begin{pmatrix}
1 &0& 0 \\
0&1&0 \\
0 &0& 1
\end{pmatrix}~~~.
\label{lddiag}
\end{align}
Using Eqs.\ (\ref{ludiag}) and (\ref{lddiag}) the CKM mixing matrix (neglecting the CP phases) in the model comes out to be
\be
\Bigl\lvert\mathcal{U}_{CKM}\Bigr\rvert\simeq
\begin{pmatrix}
0.974 &0.228& 0.002 \\
0.228&0.973&0.04\\
0.008 &0.04& 0.999
\end{pmatrix}
\thicksim
\begin{pmatrix}
1 &0& 0 \\
0&1&0 \\
0 &0& 1
\end{pmatrix}~~~.
\label{gj_ckm}
\eq
From the above prediction the quark mixing angles are 
\be
\begin{split}
&\theta^{~q}_{23}~\simeq~~2.3^{\circ}~~~,\\
&\theta^{~q}_{13}~\simeq~~0.12^{\circ}~~~,\\
&\theta^{~q}_{12}~\simeq~~13.2^{\circ}~~~.
\end{split}
\label{egjqangles}
\eq

Comparing Eqs.\ (\ref{ckmobs}) and (\ref{gj_ckm}) we see that the CKM matrix in the Georgi-Jarlskog model captures very well the features of the CKM texture. Subjecting the Yukawa coupling matrix in Eq.\ (\ref{yleptonlcal}) to bi-diagonalization gives the left diagonalizing matrix for the charged leptons as, for $\lambda\thicksim 0.23$,
 \be
\Bigl\lvert\mathcal{U}^{l\pm}_{L}\Bigr\rvert~~\simeq~~
\begin{pmatrix}
0.997& 0.076& 0 \\
0.076& 0.997& 0 \\
0 &  0& 1
\end{pmatrix}
~
\thicksim
 ~
\begin{pmatrix}
1 &0& 0 \\
0&1&0 \\
0 &0& 1
\end{pmatrix}~~~.
\label{gj_ulepL}
\eq

Thus, as seen above, the GJ textures are very attractive phenomenologically and capture to a large extent the features of the quark and charged lepton sectors. Now one may ask whether it is possible in the context of the GJ $SU(5)$ GUT and the simplest family symmetry based on $U(1)$ subfactors augmented with $\mathbb{Z}_{n}$ discrete groups to arrive naturally at the Yukawa coupling matrices in Eqs.\ (\ref{yupl})--(\ref{yleptonl}) with the least number of assumptions.

\par
It will be shown in the following  that the Georgi-Jarlskog texture is very difficult to implement with any number of $U(1)$ and $\mathbb{Z}_n$ subfactors without further assumptions. We would like to investigate the possibility of using a family symmetry ${\cal G}=U(1)_1 \times \dots U(1)_N$ to generate quark Yukawa matrices of the form in Eqs.\ (\ref{yupl}) and (\ref{ydownl}), i.e. using the FN mechanism. The fermion charges under $\cal G$ are
\be
\begin{split}
\bm{ 5}^* &: (x_i^{(1)},x_i^{(2)},\dots,x_i^{(N)})~~~,\\
\bm{ 10} &: (y_i^{(1)},y_i^{(2)},\dots,y_i^{(N)})~~~,
\end{split}
\eq
where $i=1,2,3$ labels the generation and the superscript labels the $U(1)$ in $\cal G$ to which the charge refers. Define 

\be
\begin{split}
(Y_d)_{ij}&=\log_{\lambda}\left[(\mathcal{Y}_{D})_{ij}\right]=\sum_{n=1}^N~  (x_i^{(n)}+y_j^{(n)})~~~,\\
(Y_u)_{ij}&=\log_{\lambda}\left[(\mathcal{Y}_{U})_{ij}\right]=\sum_{n=1}^N (y_i^{(n)}+y_j^{(n)})~~~.
\end{split}
\eq
Using Eqs.\ (\ref{yupl}) and (\ref{ydownl}) we get the simultaneous equations:
\begin{align}
&&&&&
&\sum_{n=1}^N  (x_1^{(n)}+y_2^{(n)})&=6~,
&\sum_{n=1}^N  (x_2^{(n)}+y_1^{(n)})&=6~,\nonumber
&&&&&&\\
&&&&&
&\sum_{n=1}^N  (x_2^{(n)}+y_2^{(n)})&=5~,
&\sum_{n=1}^N  (x_3^{(n)}+y_3^{(n)})&=3~,
&&&&&&\\
&&&&&
&\sum_{n=1}^N (y_1^{(n)}+y_2^{(n)})&=6~,
&\sum_{n=1}^N (y_2^{(n)}+y_3^{(n)})&=2~,\nonumber
&&&&&&\\
&&&&&
&\sum_{n=1}^N (y_3^{(n)}+y_3^{(n)})&=0\nonumber~~~.
\end{align}
In the process of solving the above linear equations one immediately arrives at two inconsistent algebraic equations,
\be
\sum_{n=1}^N (y_1^{(n)}-y_2^{(n)})=1\neq\sum_{n=1}^N (y_1^{(n)}-y_2^{(n)})=2~~~.\\
\eq
If we add additional $\mathbb{Z}_k$ factors to $\cal G$ the argument proceeds essentially unchanged. The charges from the cyclic subfactors modify the above linear equations in a straightforward manner; we perform the same manipulations and arrive at an inconsistency similar to that above.  We conclude that we cannot generate the Georgi-Jarlskog textures purely from a family symmetry containing only $U(1)$ and $\mathbb{Z}_k$ subfactors without resorting to further assumptions or by extending the Higgs sector. This is further backed up by numerical studies, especially if we restrict ourselves to considering charges consistent with holomorphy.
\par
An early and pioneering study using the Georgi-Jarlskog texture in $SO(10)$ was that of Harvey, Reiss and Ramond \cite{hrr}. In their model the GJ texture is implemented by three $\bm{126}$ Higgs scalars and one $\bm{ 10_{1}+i~10_{2} }$ Higgs of $SO(10)$ which have VEVs along definite directions.
\par
From the discussions in this section it is a fair assessment that incorporating finer details in the $SU(5)$ toy model would require us to extend the scalar sector or make other additional assumptions rather than just `hierarchical'  versus `democratic' texture for the mixing matrices. The Georgi-Jarlskog model is very attractive from a phenomenological point of view but imposing the texture zeroes requires some additional Higgs or other mechanisms for suppressing some entries and generating texture zeroes.
 \end{section}

\begin{section}{Conclusions}
The Yukawa coupling constants and mixing angles are among the most poorly understood features of the SM, since in the theory they are arbitrary parameters whose values are set by experiments alone. The discovery of non-zero masses for the neutrinos and a lepton mixing matrix far from the unit matrix have rekindled many studies attempting to predict fermion masses and mixing angles. Among the most interesting attempts in this regard are GUTs and family symmetries which relate some of the previously arbitrary parameters in the SM. 
\par
In this paper we have attempted to discuss the general features of GUTs with Abelian family symmetries taking the simplest GUT group $SU(5)$ as an example. One of the crucial questions that is to be understood today is why the mixing in the lepton sector is very large as compared to the quark sector. So it is interesting to explore various mechanisms that may lead to this asymmetry. We constructed a $SU(5)$ toy model with abelian family symmetries with as few assumptions as possible that replicates the observed SM mass hierarchy and mixing matrices to lowest approximation. It is seen that to include further details into the model one has to make further assumptions, add extra scalar fields or fine-tune some of the charges. Nevertheless from the point of view of incorporating the general features of the mass hierarchy and mixing angles approximately with as few assumptions as possible, the toy model has been modestly successful. The main assumptions in our model building have been the implementation of `hierarchical' quark/charged lepton sectors and a `democratic' light neutrino sector in terms of the mixing matrices. The charges of the representations as well as the VEV parameters of the flavon fields were determined purely from phenomenology. This determination of the family symmetry charges and VEV parameters give values that are of $\mathcal{O}(1)$ without any fine-tuning. 
\par
We have ignored the question of CP violating phases throughout this study. In the lepton sector apart from a normal CP phase we also expect two Majorana phases. A study of the CP phases in the CKM and PMNS matrices is especially pertinent in the context of the baryon asymmetry in the universe. It would be interesting to extend the model by incorporating possible CP phases in the quark and lepton sectors.
It would also be interesting to explore along similar lines the general features of the generation of mass hierarchy and quark/lepton mixing matrices in more interesting GUT groups such as $SO(10)$ and $E_{6}$.\\
\end{section}


\par

\begin{acknowledgments}
We thank M. Yu. Khlopov, M. Picariello and J. W. F. Valle for helpful correspondence regarding references. A.M.T. would like to acknowledge support from the Subrahmanyan Chandrasekhar Memorial Fellowship during this study. D.M. would like to thank the Robert G. Sachs Fellowship Fund for support. This work was supported in part by the United States Department of Energy under Grant No. DE-FG02-90ER40560.
\end{acknowledgments}



\end{document}